\documentclass[aps,floatfix,twocolumn,showpacs]{revtex4}
\usepackage{graphicx,amssymb,amsmath,float,braket}
\usepackage[caption=false]{subfig}

\newcommand{\Mod}[1]{\ (\text{mod}\ #1)}

\begin{document}

\title{Maximal LELM Distinguishability of Qubit and Qutrit Bell States Using Projective and Non-Projective Measurements}
\author{N. Leslie}
\email{nathaniel_leslie@berkeley.edu}
\affiliation{Department of Physics, University of California, 366 LeConte Hall, Berkeley, CA 94720, U.S.A.}
\affiliation{Department of Physics, Harvey Mudd College, 301 Platt Blvd., Claremont, CA 91711, U.S.A.}
\author{J. Devin}
\email{jdevin@stanford.edu}
\affiliation{Department of Physics, Stanford University, 382 Via Pueblo, Stanford, CA 94305}
\author{T.W. Lynn}
\email{lynn@hmc.edu}
\affiliation{Department of Physics, Harvey Mudd College, 301 Platt Blvd., Claremont, CA 91711, U.S.A.}
\date{\today }

\begin{abstract}
Numerous quantum information protocols make use of maximally entangled two-particle states, or Bell states, in which information is stored in the correlations between the two particles rather than their individual properties. Retrieving information stored in this way means distinguishing between different Bell states, yet the well known no-go theorem establishes that projective linear evolution and local measurement (LELM) detection schemes can only reliably distinguish three of the four qubit Bell states. We establish maximum distinguishability of the qutrit Bell states of bosons via projective LELM measurements; only three of the nine Bell states can be distinguished. Next, we extend to the case of non-projective measurements. We strengthen the no-go theorem by showing that general LELM measurements cannot reliably distinguish all four qubit Bell states. We also establish that at most five qutrit Bell states can be distinguished with generalized LELM measurements. 
\end{abstract}
\pacs{03.67.-a,03.67.Hk,42.50.Dv}
\maketitle

 
\section{Introduction}
\label{Section:Intro}

Bell states form a fully entangled basis for the Hilbert space of a bipartite system. Measurements in these bases are required in many applications, including quantum teleportation~\cite{TeleportationProtocol,TeleportationExperiment}, quantum repeaters~\cite{QuantumRepeaters,RepeaterImplementation}, quantum dense coding~\cite{DenseCoding}, and quantum error correction~\cite{QuantumComputing,ErrorCorrection}.  Deterministic and unambiguous discrimination between these Bell states is often useful, but it has been shown that all four qubit Bell states cannot be distinguished this way, even in principle, with a device restricted to linear evolution and local measurement (LELM) and projective measurement \cite{BellMeasurements}. Despite these limitations, LELM devices remain of significant interest in realizations of entanglement-based protocols because of their higher success probabilities overall~\cite{entswappingexpt,densecodingexpt,tpexpt,superdenseexpt,groundsatelliteTPexpt,TimeTeleport,OAMTeleport,ququarts}. LELM distinguishability limits via projective measurement are already well-established for qubit variables.  For a pair of particles entangled in a single qubit variable, at most three of the four Bell states can be reliably distinguished \cite{MethodsTeleportation, BellMeasurements}. Furthermore, for a pair of particles entangled in $n$ qubit variables, a maximum of $2^{n+1}-1$ of the $4^n$ hyper-entangled Bell states can be reliably distinguished \cite{HyperentangledDistinguishability}. However, for more general $d$-state qudit variables, distinguishability limits have not yet been established. These limits are relevant to experiments with photons entangled in variables such as orbital angular momentum (OAM) and frequency; OAM has already been used as a qutrit and qudit variable in experimental demonstrations of entanglement and quantum information protocols \cite{2002ThreeDEntanglement,2015TwistedLight}.

Previous work has assumed that optimal linear devices use projective measurements~\cite{BellMeasurements, MethodsTeleportation, WeiKwiat2007, HyperentangledDistinguishability}, since the non-deterministic nature of general POVMs suggests on the surface that they should not increase the range of reliable measurements. However, non-projective measurements are useful in many distinguishability applications; for instance, they can be used for minimum error discrimination, the task of discriminating between a set of non-orthogonal states while minimizing the probability of misidentifying the correct one \cite{Helstrom}. They also enable unambiguous state discrimination between non-orthogonal states, albeit with a non-zero probability of getting an inconclusive result \cite{MikeandIke}. Thus it is important to question whether non-projective measurements can be used to enhance deterministic and unambiguous discrimination of Bell states. In addition to the experimental applications, understanding LELM distinguishability limits may provide a deeper theoretical understanding of entanglement and quantum information resources.

Here we address several of these questions by establishing theoretical limits on LELM distinguishability of bipartite entangled states in several scenarios.  We first set out a notational framework for considering a general LELM apparatus.  Using that framework, we identify optimal LELM distinguishability for the Bell states of a pair of qutrits, using projective measurements.  Finally, we place limits on the LELM distinguishability of qubit and qutrit Bell states using generalized measurement.

\section{Notation and Background}
\label{Section:NotationandBackground}

\begin{figure}[tb]
	\centering
	\includegraphics[scale=0.25]{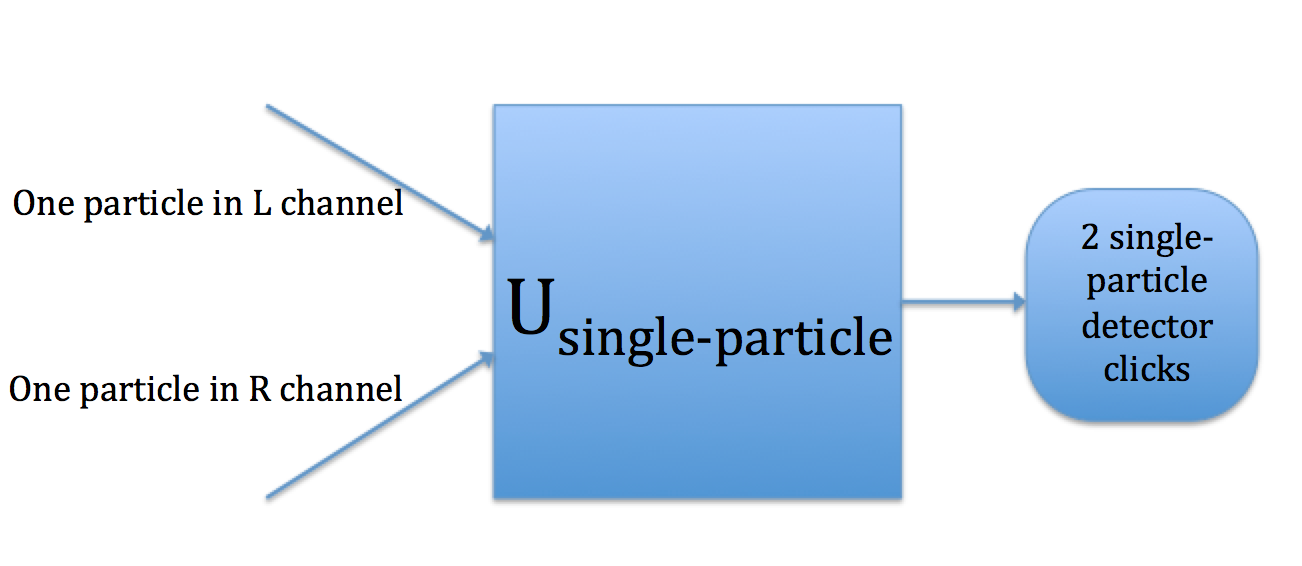}
	\caption{Two identical qudit particles enter this LELM apparatus in a symmetrized (or antisymmetrized) state with one particle in each of channels $L$ and $R$.  Both particles undergo the same single-particle unitary operation, though the single-particle unitary operates in the $2d$-dimensional space of qudit variable and channel, so it may mix the channels in the transformation from input to output modes. The outputs are detected by two detectors which, along with the single-particle unitary operator, comprise a POVM, which may be either projective or non-projective.}
	\label{fig:lelm}
\end{figure}

To establish theoretical limits on LELM Bell state discrimination, we first focus on several pieces of notation and terminology that will be useful in analyzing both the Bell states and the LELM apparatus. We consider two particles incident on a detection apparatus via two separate input channels, denoted $L$ and $R$. The fundamental restrictions we explore in this work arise from using a non-entangling apparatus -- specifically an LELM apparatus -- to distinguish between maximally-entangled states.  A general two-particle LELM apparatus is sketched in Figure \ref{fig:lelm}.  The first restriction on the LELM apparatus is that it must evolve each particle independently, although it may mix the two input channels in its transformation of input modes to output modes. The second restriction is on the measurement. Each measurement is local, corresponding to the detection of a particle in a particular output channel, so assuming perfect detector efficiency and number resolution, there will be exactly two detections. We do not consider any auxiliary modes because it has been shown that they do not improve distinguishability of any signal states in an LELM apparatus when the signal states have a fixed particle number \cite{vanLoockLut,Carollo_2002}.

We build up notation for general bipartite qudit Bell states by analogy with the more familiar qubit Bell states, shown here:

\begin{equation}
\begin{split}
\ket{\Phi^+}&=\frac{1}{\sqrt{2}}\big(\ket{0,L}\ket{0,R}+\ket{1,L}\ket{1,R}\big)\\
\ket{\Phi^-}&=\frac{1}{\sqrt{2}}\big(\ket{0,L}\ket{0,R}-\ket{1,L}\ket{1,R}\big)\\
\ket{\Psi^+}&=\frac{1}{\sqrt{2}}\big(\ket{0,L}\ket{1,R}+\ket{1,L}\ket{0,R}\big)\\
\ket{\Psi^-}&=\frac{1}{\sqrt{2}}\big(\ket{0,L}\ket{1,R}-\ket{1,L}\ket{0,R}\big).
\end{split}
\label{eq:qubitbellstates}
\end{equation}
The qubit Bell states form an entangled basis for the Hilbert space of the two-qubit system.  These Bell states are written above in terms of an unentangled basis, made from tensor products of two single-particle states in the standard basis, composed of states of definite variable values ($\{\ket{0}, \ket{1}\}$ for each qubit). We will refer to such unentangled two-particle basis states ($\{\ket{0,L}\ket{0,R}, \ket{0,L}\ket{1,R}, \ket{1,L}\ket{0,R}, \ket{1,L}\ket{1,R}\}$ in the two-qubit case) as \textit{joint-particle kets}.

Likewise, it is possible to write sets of Bell states that are bases for the state spaces of a general two-qudit ($d$-state variable) system. The qubit Bell states are labelled with $\Psi$ and $\Phi$ based on the correlation between the variable values for the two particles, and with a $+$ or $-$ based on the relative phase between the terms. For qudit Bell states, the $\Psi$/$\Phi$ distinction is generalized to a notion of \textit{correlation class}; Bell states in correlation class $c$ are made up of joint-particle kets of the form $\ket{j,L}\ket{j+c \Mod{d},R}$. The $+$/$-$ distinction is generalized to the idea of \textit{phase class}; Bell states in the same correlation class have consecutive terms in the joint-particle ket representation differing by the same phase factor. The general qudit Bell state in correlation class $c$ and phase class $p$ is given by
\begin{equation}
\ket{\Psi_c^p}=\frac{1}{\sqrt{d}}\sum_{j=0}^{d-1}e^{i2\pi pj/d}\ket{j,L}\ket{j+c \Mod{d},R}.
\label{eq:quditbellstate}
\end{equation}

As $p$ and $c$ can both range from 0 to $d-1$, we arrive at the full $d^2$ qudit Bell states.

In this paper we often consider entangled states of identical particles, as these arise most commonly in experimental scenarios and provide in general the most possibility for LELM distinguishability.  To treat bosonic or fermionic qudits properly, the Bell states must be symmetrized or antisymmetrized.  Thus the states of Equation \ref{eq:quditbellstate} are modified to become
\begin{equation}
\begin{split}
\ket{\Psi_c^p}=\frac{1}{\sqrt{2d}}\sum_{j=0}^{d-1}&e^{i2\pi pj/d}\Big(\ket{j,L}_1\ket{j+c \Mod{d},R}_2\\
 &\pm \ket{j+c \Mod{d},R}_1\ket{j,L}_2\Big),
\end{split}
\label{eq:quditbellstatesymm}
\end{equation}
where the $+$ sign applies to bosons and the $-$ sign to fermions.

As our notation above suggests, the input channel can be considered an additional aspect of the input state of each particle; in that case, there are $2d$ basis states for each particle entering the apparatus:
\begin{equation}
\{\ket{0,L},\ket{0,R},\hdots,\ket{d-1,L}, \ket{d-1,R}\}.
\label{eq:basiskets}
\end{equation}
Following the notation of \cite{HyperentangledDistinguishability}, we label these single-particle basis states $\ket{\varphi_m}, m\in\{0,\hdots,2d-1\}$, with annihilation operators $\hat{a}_m$.  We order the basis as in Equation \ref{eq:basiskets}, so that $\ket{\varphi_{2s}}=\ket{s,L}$ and $\ket{\varphi_{2s+1}}=\ket{s,R}$.  In this notation, the qudit Bell states appear as
\begin{equation}
\begin{split}
\ket{\Psi_c^p}=\frac{1}{\sqrt{2d}}\sum_{j=0}^{d-1}e^{i2\pi pj/d}\Big(&\ket{\varphi_{2j}}_1\ket{\varphi_{2(j+c \Mod{d})+1}}_2\\
 \pm &\ket{\varphi_{2(j+c \Mod{d})+1}}_1\ket{\varphi_{2j}}_2\Big).
\end{split}
\label{eq:quditbellstatesymm2}
\end{equation}
Each particle entering the system undergoes linear evolution as specified by the unitary transformation $\hat{U}_{single-particle}$ acting in the $2d$ dimensional single-particle Hilbert space of qudit variable and channel, before moving on to the detection phase. A general detection is described by a positive operator-valued measure (POVM), which we will describe in Section \ref{Section:Non-Projective}. Until then, we will consider only projective measurement, which allows significant simplifications.

One method of determining whether a set of states can be distinguishable using projective measurements is detailed in \cite{vanLoockLut}. The idea is that, to be reliably distinguishable, states must remain orthogonal after one particle has been detected in any detector. In general, we can write the annihilation operator for a detection mode as
\begin{equation}
\hat{c}=\nu_0\hat{a}_0+\nu_1\hat{a}_1+\ldots+\nu_{2d-1}\hat{a}_{2d-1},
\label{eq:annihilationoperator}
\end{equation}
where $\hat{a}_i$ is the annihilation operator for a particle in the state $\ket{\varphi_i}$. If a state $\ket{\Psi}$ has one particle detected in this detector, the remaining state is
\begin{equation}
\hat{c}\ket{\Psi}.
\label{eq:oneannihilation}
\end{equation}
For a set of states $\{\ket{\Psi_i}\}$ to be distinguishable, these remaining states must be orthogonal, giving
\begin{equation}
\bra{\Psi_k}\hat{c}^\dagger\hat{c}\ket{\Psi_l}=0 \qquad \forall k \neq l.
\label{eq:criteria}
\end{equation}
These are only necessary conditions on one detector; they are definitely not sufficient for showing that a set of states is distinguishable. For a set of states to be distinguishable, after any one detector fires, all of the states must trigger different second detectors. These conditions only establish that the states after one detection are orthogonal and hence in principle capable of triggering a non-overlapping set of second detectors.  Still, these necessary conditions are very useful for showing that sets of states are indistinguishable.

\section{Maximum Distinguishability of Qutrit Bell States with Projective Measurement}
\label{Section:Projective}
In this section we restrict our investigations to the case of local \textit{projective} measurement. In Section \ref{Section:Intro}, we listed the established limits for maximal LELM qubit distinguishability with projective measurement. However, the corresponding limit for entangled qutrits, or three-state particles, was previously unknown; here we show that for a pair of bosons entangled in a single qutrit variable, at most three of the nine Bell states can be reliably distinguished by a projective LELM apparatus.

\begin{figure}[tb]
	\centering
	\includegraphics[scale=0.25]{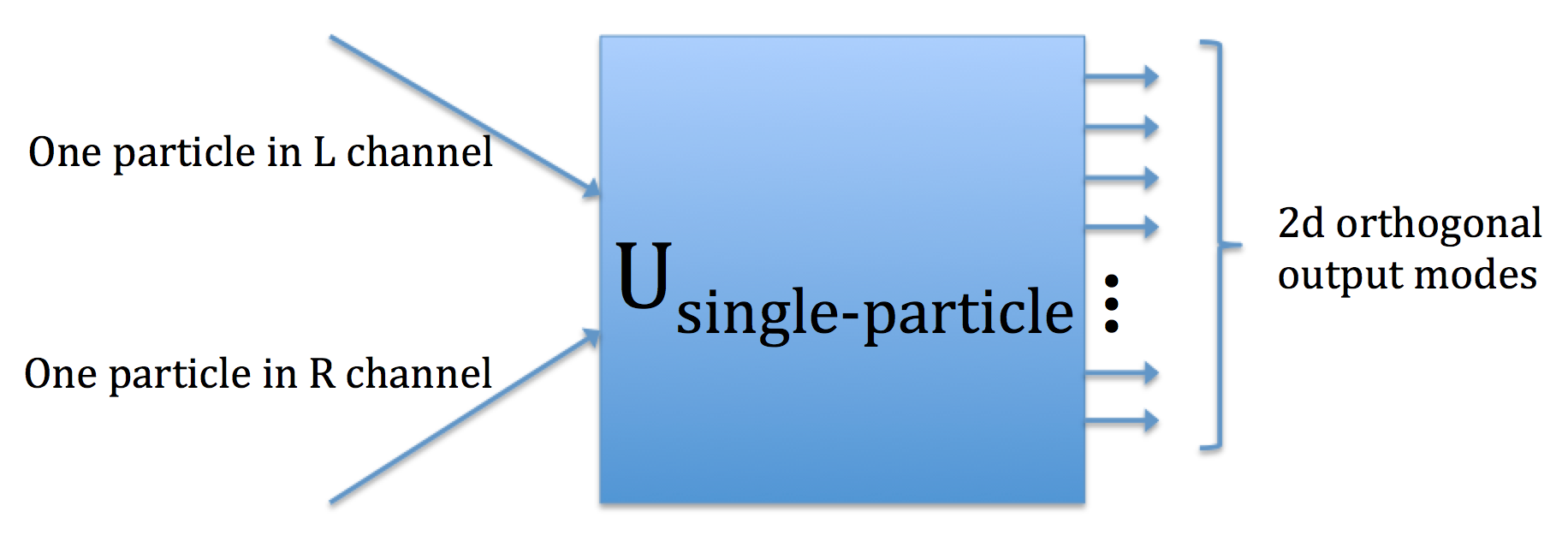}
	\caption{The projective LELM apparatus simply performs a single-particle unitary transformation on each particle before sending them to an orthogonal set of $2d$ projective detectors.  The transformation may mix left and right channels (beamsplitting), but does not contain conditional evolution. \label{fig:projlelm}}
\end{figure}

A projective LELM apparatus consists of a single-particle unitary transformation and output detectors, as described above, where each detection event annihilates a particle from a particular detector mode as discussed above.  If desired, a click in detector $i$ can be thought of as projecting the two-particle state onto a subspace where one particle is in the particular \textit{detector mode} $\ket{i}$ and then annihilating that particle; this way of thinking may be useful in generalizing to POVM measurements in Section \ref{Section:Non-Projective}. The particular apparatus and its detector modes are defined by the unitary transformation $U_{\text{single-particle}}$ and the modes that they act on, which yield the single-particle basis states of Equation \ref{eq:basiskets}:
\begin{equation}
\ket{\varphi_i}=U_{\text{single-particle}}\ket{i}
\label{eq:outputmodes}
\end{equation}
Then a measurement can be made in the standard basis, which tells us that we measured $\ket{i}$ when the detector detects $\ket{\varphi_i}$. This transformation $U_{\text{single-particle}}$ is a different way of expressing the same information that the annihilation operators in Equation \ref{eq:annihilationoperator}. Because the transformation is unitary and the output modes (the $2d$ basis kets from Equation \ref{eq:basiskets}) are orthogonal, the input modes must also be orthogonal. This apparatus is shown in Figure \ref{fig:projlelm}.

Any given Bell state fires two detectors (or one detector twice). A pair of detections is called a \textit{detection signature}. We cannot simply write the detection signatures as a raw tensor product of two detection modes $\ket{i}\otimes\ket{j}$, because that contains terms corresponding to two particles in the left input channel and to two particles in the right. Thus we introduce the projection operator $\hat{P}_{LR}$, which projects these tensor products onto the subspace of two-particle states where one comes from the left input and one comes from the right (and then renormalizes). The detection signature corresponding to a detection in detector $i$ and a detection in detector $j$ is then
\begin{equation}
\ket{i}\ket{j}=\hat{P}_{LR}\ket{i}\otimes\ket{j}.
\label{eq:detectionsignature}
\end{equation}
Respecting the statistics of the identical particles leads us to an implicit symmetrization or antisymmetrization for two-detector states as well, so that
\begin{equation}
\ket{i}\ket{j}=\frac{1}{\sqrt{2}}\big(\ket{i}_1\ket{j}_2\pm\ket{j}_1\ket{i}_2\big),
\label{eq:statistics}
\end{equation}
where the $+$ sign applies to bosons and the $-$ sign to fermions.

Two Bell states are reliably distinguishable when they give rise to a nonoverlapping set of detection signatures.  In order to determine which Bell states can lead to a particular detection signature, it will be useful to express both detection signatures and Bell states in terms of the joint-particle kets that are tensor products of one qudit state in $L$ with another qudit state in $R$.  Detection signatures can be naturally expressed this way via Equations \ref{eq:detectionsignature} and \ref{eq:outputmodes}, while qudit Bell states are already expressed this way in Equation \ref{eq:quditbellstatesymm2}.

We now focus our discussion on entangled qutrits. With implicit symmetrization or antisymmetrization from Equation \ref{eq:statistics}, the nine qutrit Bell states are given, according to Equation \ref{eq:quditbellstate}, by
\begin{equation}
\begin{split}
\ket{\Phi_0^0}=&\frac{1}{\sqrt{3}}\big(\ket{0,L}\ket{0,R}+\ket{1,L}\ket{1,R}+\ket{2,L}\ket{2,R}\big)\\
\ket{\Phi_0^1}=&\frac{1}{\sqrt{3}}\big(\ket{0,L}\ket{0,R}+e^{i\frac{2\pi}{3}}\ket{1,L}\ket{1,R}\\
&+e^{i\frac{4\pi}{3}}\ket{2,L}\ket{2,R}\big)\\
\ket{\Phi_0^2}=&\frac{1}{\sqrt{3}}\big(\ket{0,L}\ket{0,R}+e^{i\frac{4\pi}{3}}\ket{1,L}\ket{1,R}\\
&+e^{i\frac{2\pi}{3}}\ket{2,L}\ket{2,R}\big)\\
\ket{\Phi_1^0}=&\frac{1}{\sqrt{3}}\big(\ket{0,L}\ket{1,R}+\ket{1,L}\ket{2,R}+\ket{2,L}\ket{0,R}\big)\\
\ket{\Phi_1^1}=&\frac{1}{\sqrt{3}}\big(\ket{0,L}\ket{1,R}+e^{i\frac{2\pi}{3}}\ket{1,L}\ket{2,R}\\
&+e^{i\frac{4\pi}{3}}\ket{2,L}\ket{0,R}\big)\\
\ket{\Phi_1^2}=&\frac{1}{\sqrt{3}}\big(\ket{0,L}\ket{1,R}+e^{i\frac{4\pi}{3}}\ket{1,L}\ket{2,R}\\
&+e^{i\frac{2\pi}{3}}\ket{2,L}\ket{0,R}\big)\\
\ket{\Phi_2^0}=&\frac{1}{\sqrt{3}}\big(\ket{0,L}\ket{2,R}+\ket{1,L}\ket{0,R}+\ket{2,L}\ket{1,R}\big)\\
\ket{\Phi_2^1}=&\frac{1}{\sqrt{3}}\big(\ket{0,L}\ket{2,R}+e^{i\frac{2\pi}{3}}\ket{1,L}\ket{0,R}\\
&+e^{i\frac{4\pi}{3}}\ket{2,L}\ket{1,R}\big)\\
\ket{\Phi_2^2}=&\frac{1}{\sqrt{3}}\big(\ket{0,L}\ket{2,R}+e^{i\frac{4\pi}{3}}\ket{1,L}\ket{0,R}\\
&+e^{i\frac{2\pi}{3}}\ket{2,L}\ket{1,R}\big).
\end{split}
\label{eq:qutritbellstates}
\end{equation}
Here the correlation class $c$ ranges from 0 to 2, and the phase class $p$ likewise goes from 0 to 2.

To show that $n$ is the maximum number of qutrit Bell states distinguishable using an LELM apparatus, it must be demonstrated both that some set of $n$ qutrit Bell states is distinguishable and that each set of $n+1$ qutrit Bell states is not distinguishable.
 
First, we show that $n \geq 3$. If an LELM apparatus simply measures each particle in the standard or input basis (i.e., $U_{\text{single-particle}}$ is the identity), it can distinguish between three Bell states in distinct correlation classes via the difference between the measured qutrit variable values. This simple strategy identifies a lower bound of 3 for number of distinguishable qutrit Bell states. 

We next establish a rough upper bound, that no set of five qutrit Bell states is distinguishable, by testing the necessary criteria from Equation \ref{eq:criteria}; using Mathematica, we searched for solutions to the system of equations that comes from \ref{eq:criteria} for each of the 126 sets of five qutrit Bell states. Each system of equations, along with a normalization condition on the detector mode, was submitted to Mathematica's FindInstance routine, and each system gave no solutions. This tells us that $n<5$. Thus we can already say that the maximum number of distinguishable qutrit Bell states is either three or four.

The only remaining step is to investigate sets of 4 qutrit Bell states. If we find a distinguishable set, then $n=4$. If we can check all sets and rule out distinguishability for all of them, then $n=3$. We employed the same method to check for possible distinguishability of all 126 sets of four qutrit Bell states. Of these, 54 systems had no solutions and 72 systems either gave solutions or did not complete. Thus 54 sets of four Bell states were definitely indistinguishable by projective LELM, but 72 sets of four Bell states offered the possibility of distinguishability; the solutions given by FindInstance did not extend simply to successful, complete sets of six orthogonal detectors. To determine their distinguishability or lack thereof, we employed a different approach.

\subsection{Equivalence Classes of Bell States}
\label{Subsection:EquivalenceClasses}

\begin{figure}[tb]
	\centering
	\includegraphics[scale=0.35]{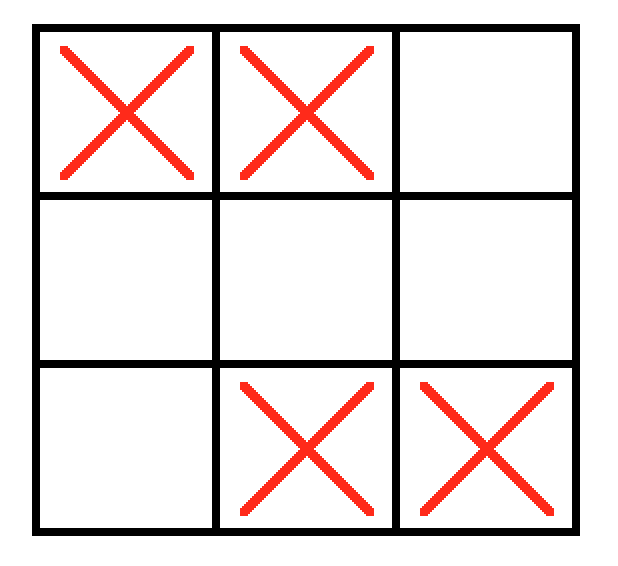}
	\caption{The tic-tac-toe diagram of the qutrit Bell-state set $\{\ket{\Psi_0^0},\ket{\Psi_0^1},\ket{\Psi_2^1},\ket{\Psi_2^2}\}$.  A Bell $\ket{\Psi_c^p}$ is placed in row $c$ according to its correlation index $c$, and in column $p$ accordindg to its phase index $p$.}
	\label{fig:00012122}
\end{figure}

\begin{figure}[tb]
	\centering
	\includegraphics[scale=0.35]{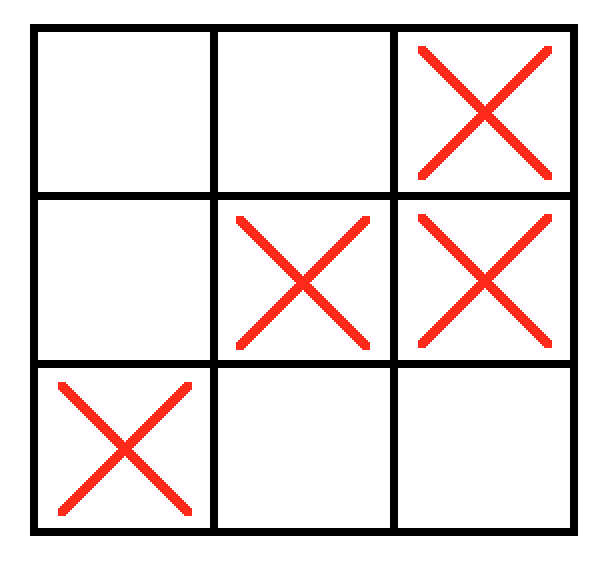}
	\caption{The tic-tac-toe diagram of the qutrit Bell-state set $\{\ket{\Psi_0^2},\ket{\Psi_1^1},\ket{\Psi_1^2},\ket{\Psi_2^0}\}$. This belongs to the tic-tac-toe winners class because it has 3 in a row diagonally.}
	\label{fig:02111220}
\end{figure}

\begin{figure}[tb]
	\centering
	\subfloat{\includegraphics[scale=0.25]{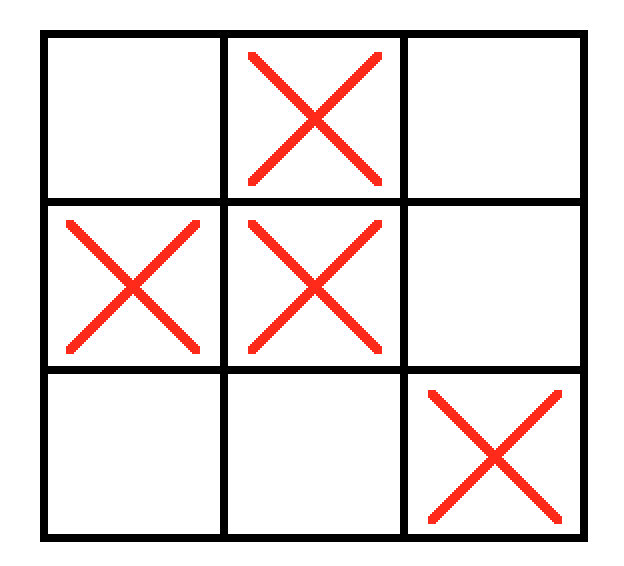}}
  	\subfloat{\includegraphics[scale=0.25]{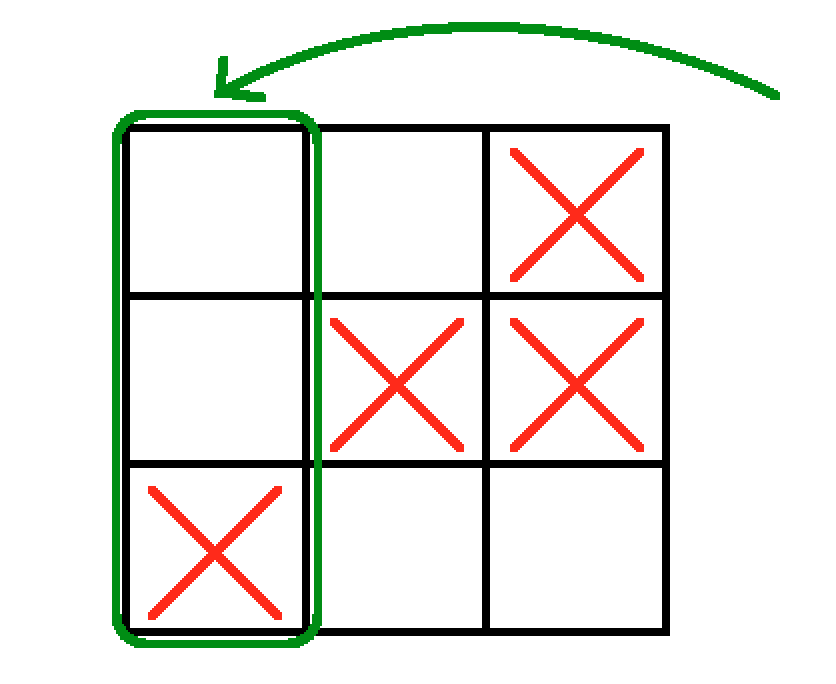}}
	\caption{The tic-tac-toe diagram of the set $\{\ket{\Psi_0^1},\ket{\Psi_1^0},\ket{\Psi_1^1},\ket{\Psi_2^2}\}$. This belongs to the tic-tac-toe winners class because it has 3 in a row diagonally when we permute the columns or allow wrap-around 3 in a row.}
	\label{fig:01101122}
\end{figure}

\begin{figure}[tb]
	\centering
	\includegraphics[scale=0.35]{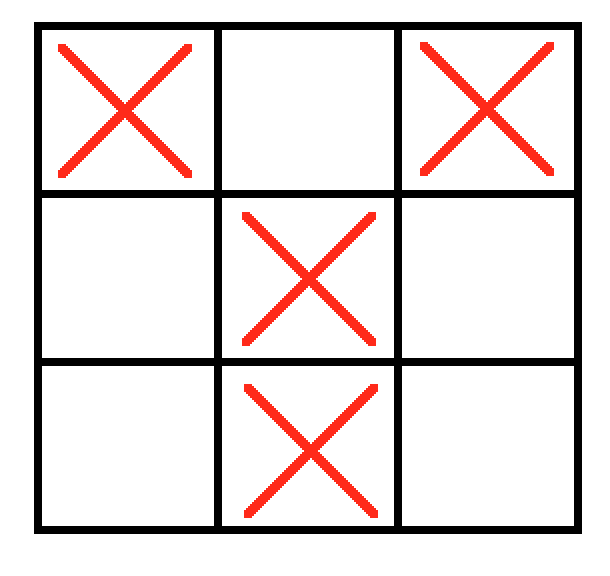}
	\caption{The tic-tac-toe diagram of the set $\{\ket{\Psi_0^0},\ket{\Psi_0^2},\ket{\Psi_1^1},\ket{\Psi_2^1}\}$. Even with column permutation or wrap-around boundaries, we cannot get 3 in a row, so it is a tic-tac-toe loser.}
	\label{fig:00021121}
\end{figure}

To understand the distinction between these two groups of sets of Bell states, it is useful to look at sets of Bell states in a $3 \times 3$ grid where the $c$ index corresponds to rows and the $p$ index corresponds to columns; we will call these grids tic-tac-toe diagrams. For example, the set $\{\ket{\Psi_0^0},\ket{\Psi_0^1},\ket{\Psi_2^1},\ket{\Psi_2^2}\}$ can be expressed by Figure \ref{fig:00012122}.
We call the 72 sets of Bell states that were inconclusive with the necessary distinguishability conditions the ``tic-tac-toe winners" class because they form boards that win at tic-tac-toe (they have 3 Xs in a row vertically, horizontally or diagonally) when we allow for column permutation or allow wrap-around boundaries. For example, the set $\{\ket{\Psi_0^2},\ket{\Psi_1^1},\ket{\Psi_1^2},\ket{\Psi_2^0}\}$ shown in Figure \ref{fig:02111220} is a tic-tac-toe winner.
The set $\{\ket{\Psi_0^1},\ket{\Psi_1^0},\ket{\Psi_1^1},\ket{\Psi_2^2}\}$ shown in Figure \ref{fig:01101122} is also a tic-tac-toe winner. But unlike $\{\ket{\Psi_0^2},\ket{\Psi_1^1},\ket{\Psi_1^2},\ket{\Psi_2^0}\}$ in Figure \ref{fig:02111220}, $\{\ket{\Psi_0^1},\ket{\Psi_1^0},\ket{\Psi_1^1},\ket{\Psi_2^2}\}$ requires either column permutation or wrap-around boundaries to give it 3 in a row.
We call the 54 sets of Bell states that were already determined to be indistinguishable the ``tic-tac-toe losers" class, because they do not form a board that wins at tic-tac-toe, even allowing column permutation or wrap-around boundaries. For example, the set $\{\ket{\Psi_0^0},\ket{\Psi_0^1},\ket{\Psi_2^1},\ket{\Psi_2^2}\}$ in Figure \ref{fig:00012122} and the set $\{\ket{\Psi_0^0},\ket{\Psi_0^2},\ket{\Psi_1^1},\ket{\Psi_2^1}\}$ in Figure \ref{fig:00021121} are both tic-tac-toe losers. 

Now we show that all of the sets of Bell states within one of these classes must share distinguishability or indistinguishability, by showing that any set of qutrit Bell states in one of these classes can be transformed into any other set in the same class using the following four operations that act on one input channel at a time:
\begin{enumerate}

\item Cycle the variable value in the right channel: 
\begin{equation}
\ket{0,R} \rightarrow \ket{1,R} \rightarrow \ket{2,R}.
\label{eq:transformation1}
\end{equation}
This increases the c index of all of the Bell states.

\item Add phases to kets in the left channel: 
\begin{equation}
\ket{1,L} \rightarrow e^{\frac{2\pi i}{3}} \ket{1,L},  \ket{2,L} \rightarrow e^{\frac{4\pi i}{3}} \ket{2,L}.
\label{eq:transformation2}
\end{equation}
This increases the p index of all of the Bell states.

\item Add phases to kets with variable value 0:
\begin{equation}
\ket{0,L} \rightarrow e^{\frac{2\pi i}{3}} \ket{0,L},  \ket{0,R} \rightarrow e^{\frac{4\pi i}{3}} \ket{0,R}.
\label{eq:transformation3}
\end{equation}
This increases the p index of the Bell states by their c index.

\item Change basis to 

\begin{subequations}
\begin{equation}
\ket{0'}=\frac{1}{\sqrt{3}}\big(\ket{0}+\ket{1}+\ket{2}\big)
\end{equation}
\begin{equation}
\ket{1'}=\frac{1}{\sqrt{3}}\big(\ket{0}+e^{\frac{2\pi i}{3}}\ket{1}+e^{\frac{4\pi i}{3}}\ket{2}\big)
\end{equation}
\begin{equation}
\ket{2'}=\frac{1}{\sqrt{3}}\big(\ket{0}+e^{\frac{4\pi i}{3}}\ket{1}+e^{\frac{2\pi i}{3}}\ket{2}\big)
\end{equation}
\label{eq:transformation4basischange}
\end{subequations}

add phases to kets in the new basis with variable value $0'$ as in transformation 3, and transform back:

\begin{subequations}
\begin{equation}
\ket{0,L} \rightarrow \frac{1}{\sqrt{3}}\big(e^{i\frac{\pi}{6}}\ket{0,L}+e^{i\frac{5\pi}{6}}\ket{1,L}+e^{i\frac{5\pi}{6}}\ket{2,L}\big),
\end{equation}
\begin{equation}
\ket{1,L} \rightarrow \frac{1}{\sqrt{3}}\big(e^{i\frac{5\pi}{6}}\ket{0,L}+e^{i\frac{\pi}{6}}\ket{1,L}+e^{i\frac{5\pi}{6}}\ket{2,L}\big),
\end{equation}
\begin{equation}
\ket{2,L} \rightarrow \frac{1}{\sqrt{3}}\big(e^{i\frac{5\pi}{6}}\ket{0,L}+e^{i\frac{5\pi}{6}}\ket{1,L}+e^{i\frac{\pi}{6}}\ket{2,L}\big),
\end{equation}
\begin{equation}
\ket{0,R} \rightarrow \frac{1}{\sqrt{3}}\big(e^{-i\frac{\pi}{6}}\ket{0,R}+e^{-i\frac{5\pi}{6}}\ket{1,R}+e^{-i\frac{5\pi}{6}}\ket{2,R}\big),
\end{equation}
\begin{equation}
\ket{1,R} \rightarrow \frac{1}{\sqrt{3}}\big(e^{-i\frac{5\pi}{6}}\ket{0,R}+e^{-i\frac{\pi}{6}}\ket{1,R}+e^{-i\frac{5\pi}{6}}\ket{2,R}\big),
\end{equation}
\begin{equation}
\ket{2,R} \rightarrow \frac{1}{\sqrt{3}}\big(e^{-i\frac{5\pi}{6}}\ket{0,R}+e^{-i\frac{5\pi}{6}}\ket{1,R}+e^{-i\frac{\pi}{6}}\ket{2,R}\big).
\end{equation}
\label{eq:transformation4}
\end{subequations}

This increases the $c$ index of the Bell states by their p index (and adds a phase of $\frac{4\pi}{3}$ if the $p$ index is nonzero).

\end{enumerate}

Originally, we used Mathematica to show that these operations transformed all sets in each class into each other. It can also be verified by hand by noting that the transformations are equivalent to the following rearrangements of the tic-tac-toe diagrams:
\begin{enumerate}

\item Cycle all rows down.

\item Cycle all columns right.

\item Cycle states in each row differently. Do nothing to the top row, cycle states in the middle row right and states in the bottom row left.

\item Cycle states in each column differently. Do nothing to the left column, cycle states in the middle column down and states in the right column up.

\end{enumerate}

Because these operations are all unitary operations on one input channel at a time, they can be realized in an LELM apparatus. If any set in a class is distinguishable, then any other set in that class could be transformed into that set within an LELM apparatus and be distinguished that way. Similarly, if one set in a class is indistinguishable, it can be transformed into any other set in its class using an LELM apparatus and so all sets in its class must be indistinguishable. 

So we must only check one set from each class to determine distinguishability of all of the qutrit Bell states. Using Mathematica, we have already determined that all of the tic-tac-toe losers are all indistinguishable separately by showing that the necessary criteria could not be satisfied, although it would have been sufficient to just show that one set did not satisfy them. Next, we only need to consider the distinguishability of any single tic-tac-toe winner to determine whether or not all of those sets are distinguishable. We found a way to rule out distinguishability for a bosonic set of qutrit Bell states belonging to the tic-tac-toe winner class, which we outline below.

\subsection{Tic-Tac-Toe Winning Sets Cannot be Distinguished}
\label{Subsection:ProofOutline}

In this section we outline a proof that, for bosonic qutrit pairs, even the tic-tac-toe winner sets of 4 Bell states cannot be reliably distinguished by projective LELM.  Details of the proof are given in Appendix A.  This proof leads to the conclusion that, at least for bosonic qutrit pairs, only three out of nine qutrit Bell states can be distinguished by LELM.

The basic strategy of the proof is to consider various possible forms that individual detector modes can take, and show that many of these forms are impossible for a detector in an apparatus that distinguishes a tic-tac-toe winning set of 4 bosonic qutrit Bell states.  Eventually enough possible forms are eliminated to demonstrate that no successful apparatus can exist. To be present in a successful apparatus, a detector mode must satisfy the necessary criteria already outlined for the 4 Bell states to be distinguished. Furthermore, no detection signature involving that mode can contain more than one of those 4 Bell states. Otherwise, such a detection would be ambiguous. Finally, we know that each one of the 9 qutrit Bell states must be present in at least one detection signature involving that mode. Since the state of any one particle in a Bell state is random, any detector can be triggered (see \cite{HyperentangledDistinguishability}). Every Bell state must be able to trigger a second detector after any first detection. These are the requirements that will allow us to eliminate various detector modes from consideration.

First, we consider a particular tic-tac-toe winning set of 4 Bell states, which we refer to as Set A.  We show that no detector mode composed exclusively of $L$ or $R$ states can be present in an apparatus to distinguish these 4 Bell states.  An apparatus distinguishing any other tic-tac-toe winner set can be turned into an apparatus distinguishing Set A via the equivalence-establishing transformations described in Section \ref{Subsection:EquivalenceClasses}, which act on one channel at a time and thus turn single-channel detector modes into single-channel detector modes.  Therefore, there cannot be single-channel detector modes in an apparatus to distinguish any tic-tac-toe winning set of 4 Bell states.  Thus no successful apparatus can include a detector mode composed of just one single-particle input state, and when considering detector modes that are superpositions of several single-particle input states, we need only consider those with contributions from both $R$ and $L$ input modes.

Next, we show that no detector mode can be a superposition of 4 single-particle input modes, by showing that the detection signature for two clicks in such a mode must include at least 8 Bell states.

Now we focus on a particular tic-tac-toe winning set of 4 Bell states, called Set B.  We show that no detector mode can be a superposition of 2 single-particle input modes (from both channels) or of 5 single-particle input modes, by showing that either would imply the existence of a detection signature containing more than one member of set B.  Further, we show that no successful apparatus can be composed entirely of detector modes that are superpositions of 3 single-particle input modes (from both channels), since at least one detection signature in such an apparatus contains multiple members of set B.

At this point, we have seen by a process of elimination that any successful apparatus to distinguish set B must contain at least one mode that is a superposition of all 6 single-particle input modes.  However, by applying the necessary criteria of Equation \ref{eq:criteria} to such a 6-ket mode, we find that it cannot be present in a successful apparatus either.  Therefore no LELM apparatus exists that can reliably distinguish the members of set B from one another.  Since all tic-tac-toe winning sets are equivalent, it follows that no tic-tac-toe winning sets of 4 bosonic qutrit Bell states are distinguishable via LELM.

Lastly, because both the tic-tac-toe winners and the tic-tac-toe losers are now all known to be indistinguishable, it is not possible for an LELM apparatus to distinguish any four out of the nine qutrit Bell states if the particles are bosons.  Hence no more than three bosonic qutrit Bell states can be distinguished by such an apparatus.  This is a notably restrictive upper limit on distinguishable states, since it is realized simply by an apparatus that measures each particle separately in the standard basis, as discussed in the beginning of Section \ref{Section:Projective}.

\section{Distinguishability of Qubit and Qutrit Bell States with General POVM Measurement}
\label{Section:Non-Projective}
So far, we have been considering distinguishability of Bell states in the special case of projective measurements. General LELM devices do not have the restriction of 2d detectors, so in principle, they offer the possibility of distinguishing more Bell states reliably. Qubit Bell state discrimination schemes using generalized measurement have been designed and experimentally implemented \cite{NonProjScheme}, but distinguishability bounds have not previously been established for this case.

A more general quantum measurement can be decribed a POVM, or positive operator-valued measure, made up of Kraus operators $\hat{E_i}$ and POVM elements $\hat{\Pi}_i=\hat{E_i}^\dagger\hat{E_i}$, which are positive operators that satisfy
\begin{equation}
\sum\hat{\Pi}_i = \hat{I},
\nonumber
\end{equation}
where $\hat{I}$ is the identity operator in the Hilbert space of the particle(s) being measured. Each one of them corresponds to a measurement that transforms a pure state
\begin{equation}
\ket{\psi} \rightarrow \frac{\hat{E_i}\ket{\psi}}{\sqrt{\bra{\psi}\hat{E_i}^\dagger\hat{E_i}\ket{\psi}}}
\label{eq:kraustransformation}
\end{equation}
with probability
\begin{equation}
p_i=\bra{\psi}\hat{E_i}^\dagger\hat{E_i}\ket{\psi}.
\label{eq:povmprob}
\end{equation}
If all of the Kraus operators are projection operators (then so are the POVM elements), the POVM is a projective measurement.  Otherwise, it is a non-projective measurement.
As mentioned in Section \ref{Section:Intro}, POVMs are already known to be useful for either making measurements with nonzero error probability or a possible inconclusive result. However, we are interested in potential use of non-projective POVMs to discriminate Bell states perfectly, with no theoretical probability of failure.

In a projective qudit Bell state discrimination scheme, the detection modes are necessarily orthogonal, which limits us to $2d$ detectors. In a general POVM scheme, the detection modes need not be orthogonal, allowing space in principle for more detectors and thus more distinct detection signatures. However, we can still argue that a general POVM is still unable to distinguish more than $2d$ Bell states.

For distinguishable particles, the two channels are automatically measured separately. We note that the state of one of the particles in a Bell state is completely random, so we cannot get any meaningful information from our first detection (see \cite{HyperentangledDistinguishability}). We would have to discriminate between the Bell states solely based on information from the second particle. In addition, even in the general POVM case, only orthogonal states can be reliably distinguished \cite{MikeandIke} (p.87). The second particle only has a $d$-state variable, so only $d$ orthogonal states exist in its Hilbert space. Therefore, only $d$ Bell states of distinguishable particles can be distinguished. 

For indistinguishable particles, the linear-evolution device can act on the $2d$-dimensional space of qudit variables and channel ($L$/$R$) variable. Again, because the state of the first particle is completely random in any Bell state, distinguishability comes only from orthogonal states of the second particle detected, but now this second particle has a maximum of $2d$ orthogonal states that might be used to distinguish the Bell states. Again, because only orthogonal states can be reliably distinguished, a general POVM cannot distinguish more than $2d$ Bell states. 

With that fact established, we can consider the maximum case of $2d$ qudit Bell states. We will show that an apparatus designed to distingush $2d$ qudit Bell states must have rank-1 Kraus operators, and we can use that fact to rule out distinguishability for $d=2$ and $d=3$.

Still, in any two-particle LELM apparatus, the two particles evolve independently through the apparatus and are individually detected, this time in a generalized measurement scheme rather than a projective one.  Without loss of generality, we can think of the ``first click" as corresponding to detection of particle $1$.  Then a ``first click" in detector $i$ represents a POVM outcome for particle $1$ and a trivial action on particle $2$.  Thus, following Equation \ref{eq:kraustransformation}, the two-particle state is transformed according to
\begin{equation}
\ket{\Psi} \rightarrow \frac{\hat{E_i}_1 \otimes \hat{I}_2\ket{\Psi}}{\sqrt{\bra{\Psi}(\hat{E_i}_1^\dagger \otimes \hat{I}_2)(\hat{E_i}_1 \otimes \hat{I}_2)\ket{\Psi}}}.
\label{eq:kraustransformationpart2}
\end{equation}
Along with this transformation, particle $1$ is annihilated in the detection event, leaving only particle $2$ to be subsequently detected. To distinguish $2d$ Bell states reliably, the remaining  particle-$2$ states from those $2d$ original Bell states must be mutually orthogonal. 

If any particle-$2$ state after the first click is mixed, we can express it as an ensemble mixture of at least two pure states. Then it is impossible to have $2d-1$ other remaining states orthogonal to all of the states in the ensemble, so it would be impossible to reliably distinguish all $2d$ remaning states. By this logic, to be able to distinguish $2d$ qudit Bell states, the particle-$2$ states after the first detection must all be pure. This means that, for every Bell state we are trying to distinguish, the states of particles $1$ and $2$ after the transformation of Equation \ref{eq:kraustransformationpart2} must be unentangled.  

Applying this transformation to a Bell state of the form of Equation \ref{eq:quditbellstatesymm2}, we get, after a first click in detector $i$,
\begin{equation}
\begin{split}
\ket{\Psi_c^p}_i=\frac{1}{\sqrt{2d}}\sum_{j=0}^{d-1}e^{i2\pi pj/d}\Big(&\hat{E_i}\ket{\varphi_{2j-1}}_1\ket{\varphi_{2(j+c \Mod{d})}}_2\\
 + &\hat{E_i}\ket{\varphi_{2(j+c \Mod{d})}}_1\ket{\varphi_{j-1}}_2\Big),
\end{split}
\label{eq:genbellstatekraus}
\end{equation}
up to an overall normalization factor which is unimportant in the argument that follows. In order for the state in Equation \ref{eq:genbellstatekraus} to be unentangled or separable into two single-particle states, all of the states $\hat{E_i}\ket{...}_1$ appearing in Equation \ref{eq:genbellstatekraus} must be scalar multiples of each other. So the states to which the Kraus operator $\hat{E}_i$ takes each particle-$1$ basis state must all be scalar multiples of some fixed state.  This means that each Kraus operator $\hat{E}_i$ must be rank 1.
Because of this, we can restrict a general Kraus operator in an apparatus that can potentially distinguish $2d$ Bell states in the following form:
\begin{equation}
\hat{E_i}=
\begin{bmatrix}
    \alpha_1n_1 & \alpha_2n_1 & \hdots & \alpha_{2d}n_1 \\
    \alpha_1n_2 & \alpha_2n_2 & \hdots & \alpha_{2d}n_2 \\
    \vdots & \vdots & \ddots & \vdots \\
    \alpha_1n_{2d} & \alpha_2n_{2d} & \hdots & \alpha_{2d}n_{2d}
\end{bmatrix}
\label{eq:genkrausop}
\end{equation}
We can now use this restriction to rule out the distiguishability of $2d$ qudit Bell states. Any apparatus that could potentially distinguish any set of $2d$ qudit Bell states would need Kraus operators of this form. We can apply such a Kraus operator to the state of the first particle that is detected in the $2d$ Bell states that we wish to distinguish to get the resulting two-particle states after the detection. This is done by applying Equation \ref{eq:kraustransformationpart2}. As we did this in a way that leaves particles 1 and 2 unentangled, the state can be expressed as a tensor product of two single-particle states. Then we annihilate the particle that we applied the Kraus operator to in order to get the remaining single-particle states to be distinguished. If we show that the resulting states are not orthogonal, they are not distinguishable with the remaining detection, which means that the Bell states that we started with are not distinguishable. We will do this for the one set of 4 qubit Bell states and all possible sets of 6 qutrit Bell states in the following subsections.

\subsection{POVM Limits for Qubit Bell States}
\label{Subsection:QubitPOVMLimits}

When we symmetrize or antisymmetrize the qubit Bell states as shown in Equation \ref{eq:quditbellstatesymm2}, we get:
\begin{subequations}
\begin{equation}
\begin{split}
\ket{\Phi^{+}}=\frac{1}{2}(\ket{0,L}\ket{0,R}\pm\ket{0,R}\ket{0,L}\\
+\ket{1,L}\ket{1,R}\pm\ket{1,R}\ket{1,L})
\end{split}
\end{equation}    
\begin{equation}
\begin{split}
\ket{\Phi^{-}}=\frac{1}{2}(\ket{0,L}\ket{0,R}\pm\ket{0,R}\ket{0,L}\\
-\ket{1,L}\ket{1,R}\mp\ket{1,R}\ket{1,L})
\end{split}
\end{equation}
\begin{equation}
\begin{split}
\ket{\Psi^{+}}=\frac{1}{2}(\ket{0,L}\ket{1,R}\pm\ket{0,R}\ket{1,L}\\
+\ket{1,L}\ket{0,R}\pm\ket{1,R}\ket{0,L})
\end{split}
\end{equation}
\begin{equation}
\begin{split}
\ket{\Psi^{-}}=\frac{1}{2}(\ket{0,L}\ket{1,R}\mp\ket{0,R}\ket{1,L}\\
-\ket{1,L}\ket{0,R}\pm\ket{1,R}\ket{0,L}).
\end{split}
\end{equation}
\label{eq:qubitbellsym}
\end{subequations}
Now we can use the $4\times4$ version of Equation \ref{eq:genkrausop} on the first particle in each of these Bell states, leaving the second particles in these states (we neglect normalization):
\begin{subequations}
\begin{equation}
\begin{split}
\ket{\Phi^{+}}_{i,2}=\frac{1}{2}(\pm\alpha_2\ket{0,L}+\alpha_1\ket{0,R}\\
\pm\alpha_4\ket{1,L}+\alpha_3\ket{1,R})
\end{split}
\end{equation}    
\begin{equation}
\begin{split}
\ket{\Phi^{-}}_{i,2}=\frac{1}{2}(\pm\alpha_2\ket{0,L}+\alpha_1\ket{0,R}\\
\mp\alpha_4\ket{1,L}-\alpha_3\ket{1,R})
\end{split}
\end{equation}
\begin{equation}
\begin{split}
\ket{\Psi^{+}}_{i,2}=\frac{1}{2}(\pm\alpha_4\ket{0,L}+\alpha_3\ket{0,R}\\
\pm\alpha_2\ket{1,L}+\alpha_1\ket{1,R})
\end{split}
\end{equation}
\begin{equation}
\begin{split}
\ket{\Psi^{-}}_{i,2}=\frac{1}{2}(\pm\alpha_4\ket{0,L}-\alpha_3\ket{0,R}\\
\mp\alpha_2\ket{1,L}+\alpha_1\ket{1,R}).
\end{split}
\end{equation}
\label{eq:qubitparticle2}
\end{subequations}
All these states must be orthogonal for the four Bell states to be distinguishable. In both the boson and fermion cases, it is straightforward to check that the six pairwise orthogonality conditions are not satisfiable. So there is no way for such an apparatus to distinguish all four qubit Bell states.

\subsection{POVM Limits for Qutrit Bell States}
\label{Subsection:QutritPOVMLimits}

\begin{figure}[tb]
	\centering
	\includegraphics[scale=0.2]{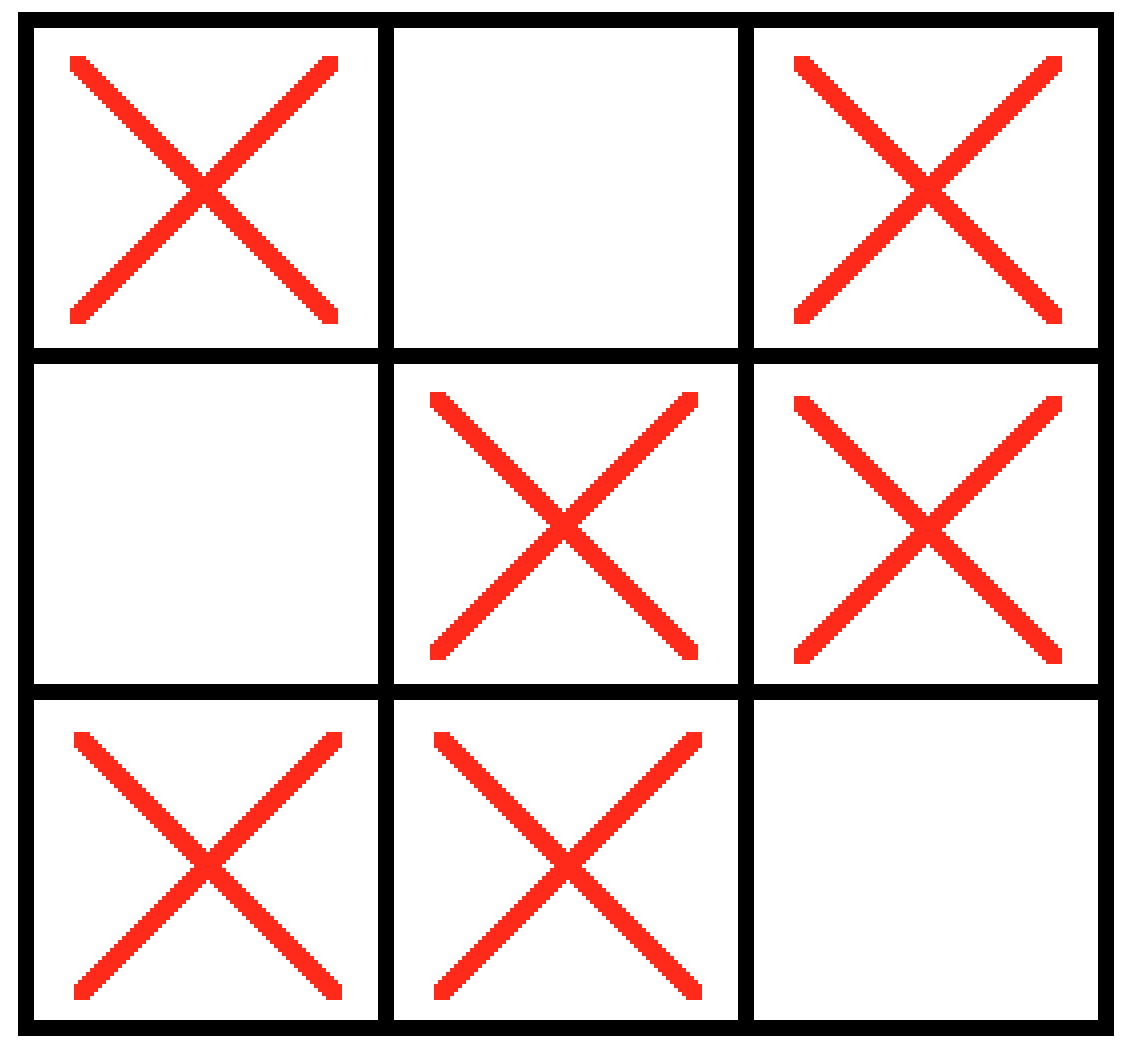}
	\caption{The tic-tac-toe diagram of the set $\{\ket{\Psi_0^0},\ket{\Psi_0^2},\ket{\Psi_1^1},\ket{\Psi_1^2},\ket{\Psi_2^0},\ket{\Psi_2^1}\}$. This belongs to the tic-tac-toe anti-winners class because with wrap-around boundary conditions or column permutation, the missing states win at tic-tac-toe.}
	\label{fig:000211122021}
\end{figure}

When we symmetrize or antisymmetrize the qutrit Bell states, we get:
\begin{subequations}
\begin{equation}
\begin{split}
\ket{\Psi_0^0}=&\frac{1}{\sqrt{6}}(\ket{0,L}\ket{0,R}\pm\ket{0,R}\ket{0,L}\\
&+\ket{1,L}\ket{1,R}\pm\ket{1,R}\ket{1,L}\\
&+\ket{2,L}\ket{2,R}\pm\ket{2,R}\ket{2,L})
\end{split}
\end{equation}    
\begin{equation}
\begin{split}
\ket{\Psi_0^1}=&\frac{1}{\sqrt{6}}(\ket{0,L}\ket{0,R}\pm\ket{0,R}\ket{0,L}\\
&+e^{i\frac{2\pi}{3}}\ket{1,L}\ket{1,R}\pm e^{i\frac{2\pi}{3}}\ket{1,R}\ket{1,L}\\
&+e^{i\frac{4\pi}{3}}\ket{2,L}\ket{2,R}\pm e^{i\frac{4\pi}{3}}\ket{2,R}\ket{2,L})
\end{split}
\end{equation}
\begin{equation}
\begin{split}
\ket{\Psi_0^2}=&\frac{1}{\sqrt{6}}(\ket{0,L}\ket{0,R}\pm\ket{0,R}\ket{0,L}\\
&+e^{i\frac{4\pi}{3}}\ket{1,L}\ket{1,R}\pm e^{i\frac{4\pi}{3}}\ket{1,R}\ket{1,L}\\
&+e^{i\frac{2\pi}{3}}\ket{2,L}\ket{2,R}\pm e^{i\frac{2\pi}{3}}\ket{2,R}\ket{2,L})
\end{split}
\end{equation}
\begin{equation}
\begin{split}
\ket{\Psi_1^0}=&\frac{1}{\sqrt{6}}(\ket{0,L}\ket{1,R}\pm\ket{1,R}\ket{0,L}\\
&+\ket{1,L}\ket{2,R}\pm\ket{2,R}\ket{1,L}\\
&+\ket{2,L}\ket{0,R}\pm\ket{0,R}\ket{2,L})
\end{split}
\end{equation}
\begin{equation}
\begin{split}
\ket{\Psi_1^1}=&\frac{1}{\sqrt{6}}(\ket{0,L}\ket{1,R}\pm\ket{1,R}\ket{0,L}\\
&+e^{i\frac{2\pi}{3}}\ket{1,L}\ket{2,R}\pm e^{i\frac{2\pi}{3}}\ket{2,R}\ket{1,L}\\
&+e^{i\frac{4\pi}{3}}\ket{2,L}\ket{0,R}\pm e^{i\frac{4\pi}{3}}\ket{0,R}\ket{2,L})
\end{split}
\end{equation}
\begin{equation}
\begin{split}
\ket{\Psi_1^2}=&\frac{1}{\sqrt{6}}(\ket{0,L}\ket{1,R}\pm\ket{1,R}\ket{0,L}\\
&+e^{i\frac{4\pi}{3}}\ket{1,L}\ket{2,R}\pm e^{i\frac{4\pi}{3}}\ket{2,R}\ket{1,L}\\
&+e^{i\frac{2\pi}{3}}\ket{2,L}\ket{0,R}\pm e^{i\frac{2\pi}{3}}\ket{0,R}\ket{2,L})
\end{split}
\end{equation}
\begin{equation}
\begin{split}
\ket{\Psi_2^0}=&\frac{1}{\sqrt{6}}(\ket{0,L}\ket{2,R}\pm\ket{2,R}\ket{0,L}\\
&+\ket{1,L}\ket{0,R}\pm\ket{0,R}\ket{1,L}\\
&+\ket{2,L}\ket{1,R}\pm\ket{1,R}\ket{2,L})
\end{split}
\end{equation}
\begin{equation}
\begin{split}
\ket{\Psi_2^1}=&\frac{1}{\sqrt{6}}(\ket{0,L}\ket{2,R}\pm\ket{2,R}\ket{0,L}\\
&+e^{i\frac{2\pi}{3}}\ket{1,L}\ket{0,R}\pm e^{i\frac{2\pi}{3}}\ket{0,R}\ket{1,L}\\
&+e^{i\frac{4\pi}{3}}\ket{2,L}\ket{1,R}\pm e^{i\frac{4\pi}{3}}\ket{1,R}\ket{2,L})
\end{split}
\end{equation}
\begin{equation}
\begin{split}
\ket{\Psi_2^2}=&\frac{1}{\sqrt{6}}(\ket{0,L}\ket{2,R}\pm\ket{2,R}\ket{0,L}\\
&+e^{i\frac{4\pi}{3}}\ket{1,L}\ket{0,R}\pm e^{i\frac{4\pi}{3}}\ket{0,R}\ket{1,L}\\
&+e^{i\frac{2\pi}{3}}\ket{2,L}\ket{1,R}\pm e^{i\frac{2\pi}{3}}\ket{1,R}\ket{2,L}).
\end{split}
\end{equation}
\label{eq:qutritbellsym} 
\end{subequations}
Now we can use the $6\times6$ version of Equation \ref{eq:genkrausop} on the first particle in each of these Bell states, which leaves the second particles in these states (again we neglect normalization):
\begin{subequations}
\begin{equation}
\begin{split}
\ket{\Psi_0^0}_{i,2}=&\frac{1}{\sqrt{6}}(\pm\alpha_2\ket{0,L}+\alpha_1\ket{0,R}\\
&\pm\alpha_4\ket{1,L}+\alpha_3\ket{1,R}\\
&\pm\alpha_6\ket{2,L}+\alpha_5\ket{2,R})
\label{eq:firstqutritparticle2}
\end{split}
\end{equation}    
\begin{equation}
\begin{split}
\ket{\Psi_0^1}_{i,2}=&\frac{1}{\sqrt{6}}(\pm\alpha_2\ket{0,L}+\alpha_1\ket{0,R}\\
&\pm e^{i\frac{2\pi}{3}}\alpha_4\ket{1,L}+e^{i\frac{2\pi}{3}}\alpha_3\ket{1,R}\\
&\pm e^{i\frac{4\pi}{3}}\alpha_6\ket{2,L}+e^{i\frac{4\pi}{3}}\alpha_5\ket{2,R})
\end{split}
\end{equation}
\begin{equation}
\begin{split}
\ket{\Psi_0^2}_{i,2}=&\frac{1}{\sqrt{6}}(\pm\alpha_2\ket{0,L}+\alpha_1\ket{0,R}\\
&\pm e^{i\frac{4\pi}{3}}\alpha_4\ket{1,L}+e^{i\frac{4\pi}{3}}\alpha_3\ket{1,R}\\
&\pm e^{i\frac{2\pi}{3}}\alpha_6\ket{2,L}+e^{i\frac{2\pi}{3}}\alpha_5\ket{2,R})
\end{split}
\end{equation}
\begin{equation}
\begin{split}
\ket{\Psi_1^0}_{i,2}=&\frac{1}{\sqrt{6}}(\pm\alpha_4\ket{0,L}+\alpha_5\ket{0,R}\\
&\pm\alpha_6\ket{1,L}+\alpha_1\ket{1,R}\\
&\pm\alpha_2\ket{2,L}+\alpha_3\ket{2,R})
\end{split}
\end{equation}
\begin{equation}
\begin{split}
\ket{\Psi_1^1}_{i,2}=&\frac{1}{\sqrt{6}}(\pm\alpha_4\ket{0,L}+e^{i\frac{4\pi}{3}}\alpha_5\ket{0,R}\\
&\pm e^{i\frac{2\pi}{3}}\alpha_5\ket{1,L}+\alpha_1\ket{1,R}\\
&\pm e^{i\frac{4\pi}{3}}\alpha_2\ket{2,L}+e^{i\frac{2\pi}{3}}\alpha_3\ket{2,R})
\label{eq:fifthqutritparticle2}
\end{split}
\end{equation}
\begin{equation}
\begin{split}
\ket{\Psi_1^2}_{i,2}=&\frac{1}{\sqrt{6}}(\pm\alpha_4\ket{0,L}+e^{i\frac{2\pi}{3}}\alpha_5\ket{0,R}\\
&\pm e^{i\frac{4\pi}{3}}\alpha_6\ket{1,L}+\alpha_1\ket{1,R}\\
&\pm e^{i\frac{2\pi}{3}}\alpha_2\ket{2,L}+e^{i\frac{4\pi}{3}}\alpha_3\ket{2,R})
\end{split}
\end{equation}
\begin{equation}
\begin{split}
\ket{\Psi_2^0}_{i,2}=&\frac{1}{\sqrt{6}}(\pm\alpha_6\ket{0,L}+\alpha_3\ket{0,R}\\
&\pm\alpha_2\ket{1,L}+\alpha_5\ket{1,R}\\
&\pm\alpha_4\ket{2,L}+\alpha_1\ket{2,R})
\end{split}
\end{equation}
\begin{equation}
\begin{split}
\ket{\Psi_2^1}_{i,2}=&\frac{1}{\sqrt{6}}(\pm\alpha_6\ket{0,L}+e^{i\frac{2\pi}{3}}\alpha_3\ket{0,R}\\
&\pm e^{i\frac{2\pi}{3}}\alpha_2\ket{1,L}+e^{i\frac{4\pi}{3}}\alpha_5\ket{1,R}\\
&\pm e^{i\frac{4\pi}{3}}\alpha_4\ket{2,L}+\alpha_1\ket{2,R})
\end{split}
\end{equation}
\begin{equation}
\begin{split}
\ket{\Psi_2^2}_{i,2}=&\frac{1}{\sqrt{6}}(\pm\alpha_6\ket{0,L}+e^{i\frac{4\pi}{3}}\alpha_3\ket{0,R}\\
&\pm e^{i\frac{4\pi}{3}}\alpha_2\ket{1,L}+e^{i\frac{2\pi}{3}}\alpha_5\ket{1,R}\\
&\pm e^{i\frac{2\pi}{3}}\alpha_4\ket{2,L}+\alpha_1\ket{2,R}).
\end{split}
\end{equation}
\label{eq:qutritparticle2} 
\end{subequations}
Now, we are left with $\binom{9}{6}=84$ sets of 6 Bell states to check for distinguishability. We can simplify this greatly by considering equivalence classes of 6-state sets that we can generate using the operations from Section \ref{Subsection:EquivalenceClasses}.
It can be verified by hand using the transformations on a tic-tac-toe diagram that these operations establish 2 classes, which we will call the tic-tac-toe anti-winners and the tic-tac-toe anti-losers. Any set of 6 states leaves out 3 of the qutrit Bell states. The location of those states is what determines which class a set belongs to.
If the missing states in a set win at tic-tac-toe with wrap-around boundary conditions or column permutation, then that set is a tic-tac-toe anti-winner. An example is shown in Figure \ref{fig:000211122021}.

\begin{figure}[tb]
	\centering
	\includegraphics[scale=0.2]{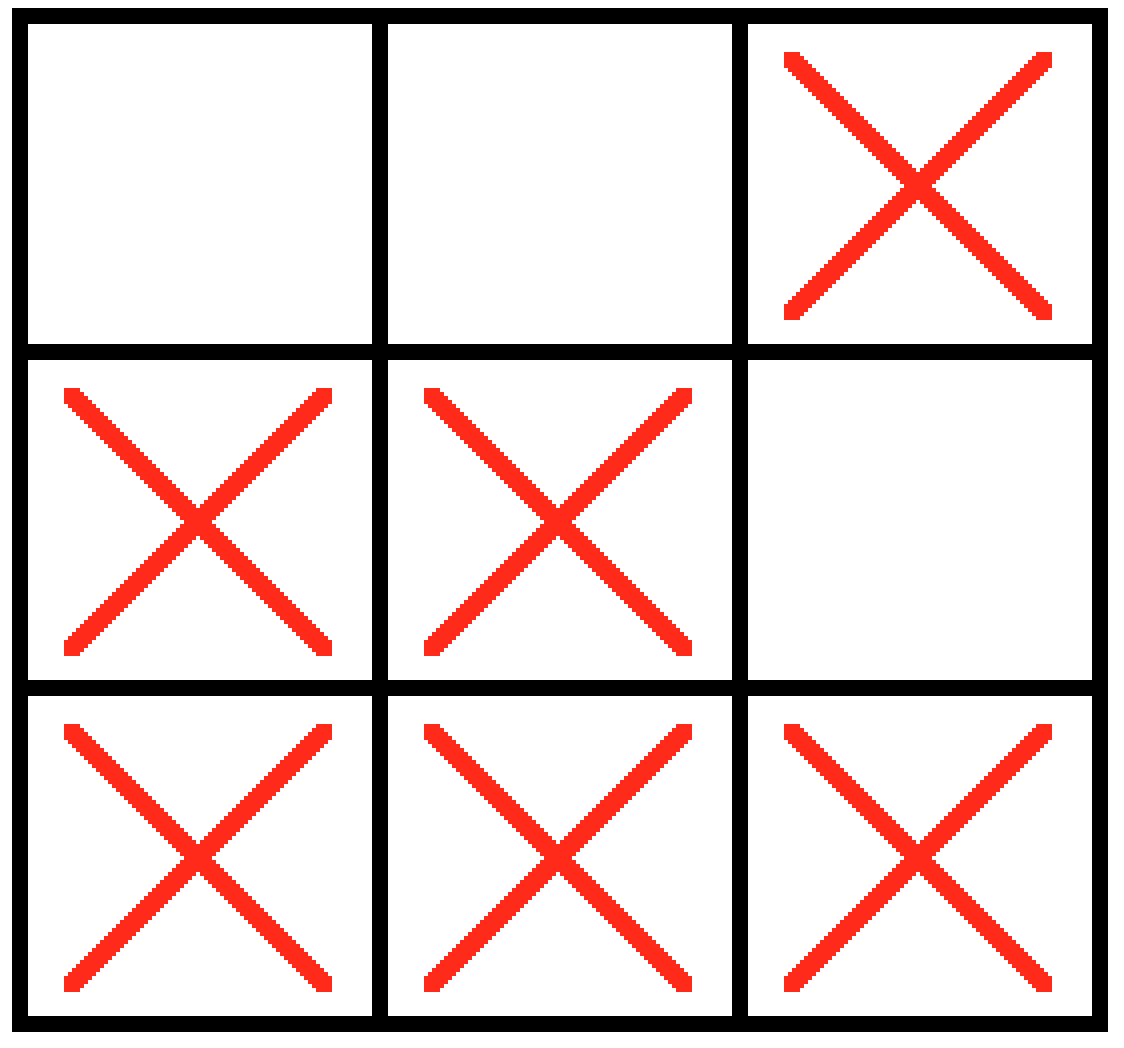}
	\caption{The tic-tac-toe diagram of the set $\{\ket{\Psi_0^2},\ket{\Psi_1^0},\ket{\Psi_1^1},\ket{\Psi_2^0},\ket{\Psi_2^1},\ket{\Psi_2^2}\}$. This belongs to the tic-tac-toe anti-losers class because with even with wrap-around boundary conditions or column permutation, the missing states do not win at tic-tac-toe.}
	\label{fig:021011202122}
\end{figure}

If the missing states in a set do not win at tic-tac-toe with wrap-around boundary conditions or column permutation, then that set is a tic-tac-toe anti-loser. An example is shown in Figure \ref{fig:021011202122}.

\begin{figure}[tb]
	\centering
	\includegraphics[scale=0.13]{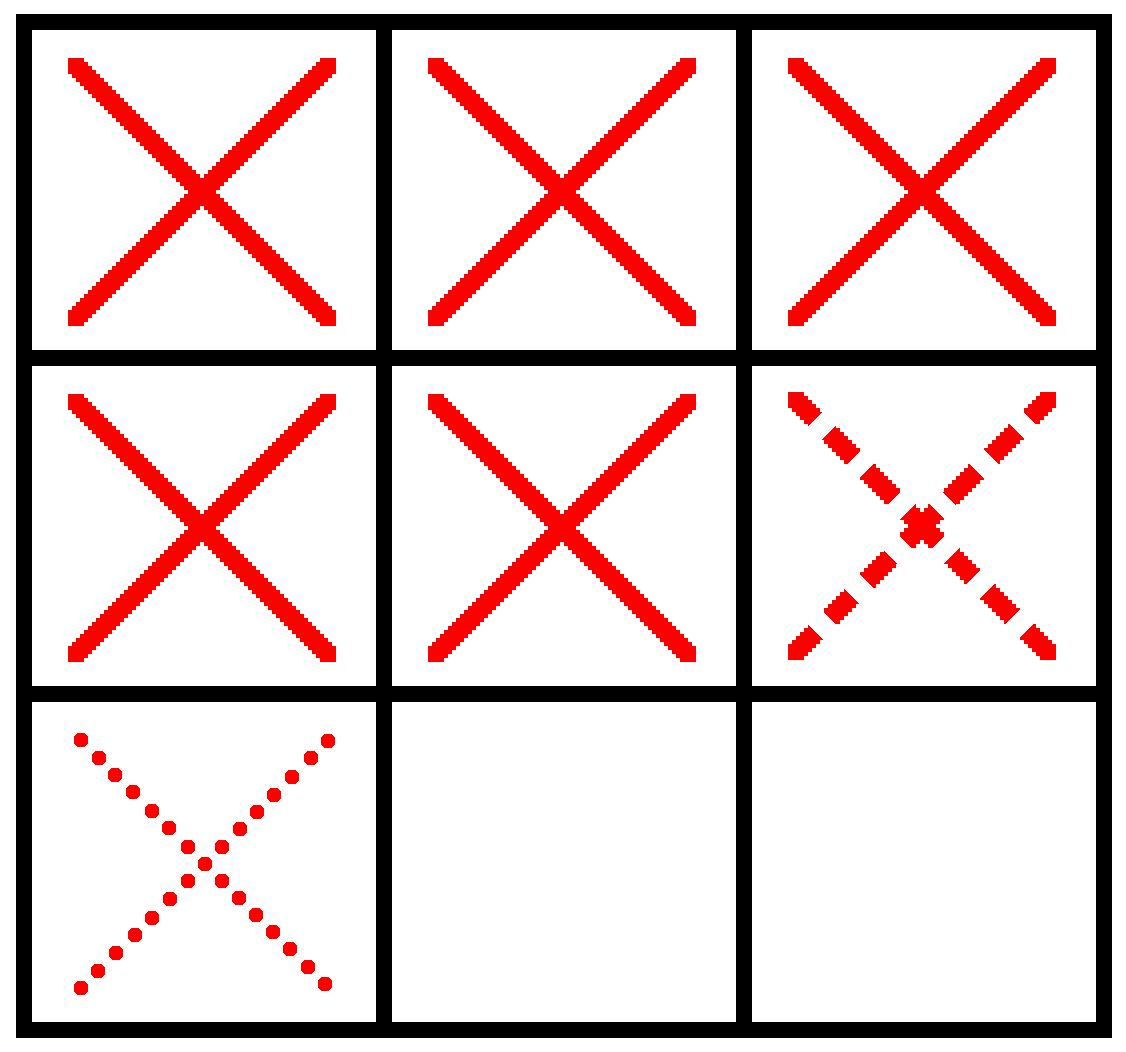}
	\caption{The subset $\{\ket{\Psi_0^0},\ket{\Psi_0^1},\ket{\Psi_0^2},\ket{\Psi_1^0},\ket{\Psi_1^1}\}$ is shown with regular Xs. The tic-tac-toe anti-winner adds $\ket{\Psi_1^2}$, shown as a dashed X. The tic-tac-toe anti-loser adds $\ket{\Psi_2^0}$, shown as a dotted X. By showing that the subset cannot remain orthogonal after one detection, we will show that both representatives from the two classes cannot remain orthogonal after one detection.}
	\label{fig:combottt}
\end{figure}

We can simultaneously show that both a representative of the tic-tac-toe anti-winners and a representative of the tic-tac-toe anti-losers cannot be distinguishable by showing that a subset of both representatives cannot be orthogonal after 1 detection. This subset is $\{\ket{\Psi_0^0},\ket{\Psi_0^1},\ket{\Psi_0^2},\ket{\Psi_1^0},\ket{\Psi_1^1}\}$ and it is a subset of the tic-tac-toe anti winner $\{\ket{\Psi_0^0},\ket{\Psi_0^1},\ket{\Psi_0^2},\ket{\Psi_1^0},\ket{\Psi_1^1},\ket{\Psi_1^2}\}$ and the tic-tac-toe anti-loser $\{\ket{\Psi_0^0},\ket{\Psi_0^1},\ket{\Psi_0^2},\ket{\Psi_1^0},\ket{\Psi_1^1},\ket{\Psi_2^0}\}$. All of these sets are shown in Figure \ref{fig:combottt}.
To show that the subset cannot remain orthogonal after one detection, we will use the general forms of the second particles in Equations \ref{eq:firstqutritparticle2}-\ref{eq:fifthqutritparticle2}. Imposing $\bra{\Psi_0^0}_{i,2}\ket{\Psi_0^1}_{i,2}=0$ and $\bra{\Psi_1^0}_{i,2}\ket{\Psi_1^1}_{i,2}=0$ (in either the fermion or boson case) gives the conditions
\begin{equation}
\big|\alpha_1\big|=\big|\alpha_3\big|=\big|\alpha_5\big| \quad \text{and} \quad \big|\alpha_2\big|=\big|\alpha_4\big|=\big|\alpha_6\big|.
\nonumber
\end{equation}
Imposing $\bra{\Psi_0^0}_{i,2}\ket{\Psi_1^0}_{i,2}=0$, $\bra{\Psi_0^1}_{i,2}\ket{\Psi_1^0}_{i,2}=0$ and $\bra{\Psi_0^2}_{i,2}\ket{\Psi_1^0}_{i,2}=0$ (again in either the fermion or boson case) gives the conditions
\begin{equation}
\alpha_2\alpha_4^*=-\alpha_1\alpha_5^*
\nonumber
\end{equation}
\begin{equation}
\alpha_4\alpha_6^*=-\alpha_3\alpha_1^*
\nonumber
\end{equation}
\begin{equation}
\alpha_6\alpha_2^*=-\alpha_5\alpha_3^*.
\nonumber
\end{equation}
Multiplying all of these together gives
\begin{equation}
\big|\alpha_2\big|^2\big|\alpha_4\big|^2\big|\alpha_6\big|^2=-\big|\alpha_1\big|^2\big|\alpha_3\big|^2\big|\alpha_5\big|^2.
\nonumber
\end{equation}
The only way that all of these conditions can be satisfied is if all of the coefficients are zero, corresponding to a null detection channel. Thus no set of 6 qutrit Bell states are distinguishable with any LELM apparatus.
\\
\section{Conclusion}

We have established distinguishability limits for qutrit Bell states with projective measurements and both qubit and qutrit Bell states with non-projective measurements. 

We have found that projective measurements can distinguish at most 3 of the 9 qutrit Bell states of identical bosons. This is a very restrictive limit, implying that in applications relying on projective LELM state distinguishability, qutrit Bell states of bosons can convey no more information than qubit Bell states.

We have strengthened the no-go theorem for qubit Bell states by showing that any LELM apparatus, using not only projective measurement but general POVMs, can only distinguish 3 out of 4 qubit Bell states. This establishes that the procedure proposed and implemented in \cite{NonProjScheme} is optimal for reliable Bell state discrimination with a POVM. We have also found that general POVMs can distinguish 5 or fewer qutrit Bell states. Finally, we have shown that, for general entangled pairs of qudits, no more than $2d$ qudit Bell states can ever be distinguished reliably with a projective or POVM LELM apparatus. All of the general POVM restrictions apply to both fermions and bosons. The bounds for non-projective measurement may in fact be identical to those for projective measurement, but that has not yet been established conclusively.

In the future, we hope to exactly determine general maximum distinguishability for qutrit Bell states with an LELM apparatus. In addition, we hope to determine limits on distinguishability of hyperentangled generalized Bell states, particularly in the experimentally relevant qubit/qutrit case.

\section{Acknowledgments}
The authors thank Neal Pisenti, Lucas Brady, Andrew Turner and Victor Shang for early contributions to this work.  N. Leslie thanks Darryl Yong for his assistance and encouragement.

\section{Appendix A}

This appendix provides details of the proof outlined in Section \ref{Subsection:ProofOutline}, showing that tic-tac-toe winning sets of bosonic qutrit Bell states are not distinguishable via a projective LELM apparatus. As we are dealing with bosons, we will be assuming the explicit symmetrization from Equation \ref{eq:statistics}, with the plus sign.

In order to show that we cannot distinguish tic-tac-toe winning sets of bosonic qutrit Bell states, we will rule out every possible detector mode in a possible apparatus. We will do this by demonstrating that detection signatures containing those modes that must exist in a complete apparatus cannot distinguish a set of tic-tac-toe winners.

If we want to distinguish two Bell states $\ket{\Psi_1}$ and $\ket{\Psi_2}$, then a detection signature $\ket{i}\ket{j}$ is not allowed if both $\bra{\Psi_1}(\ket{i}\ket{j})\neq 0$ and $\bra{\Psi_2}(\ket{i}\ket{j})\neq 0$. This would mean that both detectors could fire in the presence of both Bell states, meaning that such a detection would not be able to distinguish the two. We can see this quite easily by considering expressions of the detection signatures in the Bell basis; a viable detection signature must not have multiple Bell states that must be distinguished in its Bell-basis representation.

We see that both joint-particle states and Bell states are grouped by correlation class, so we can separate the joint-particle states into correlation classes and consider detection signatures that contain different numbers of Bell states in the same correlation class. The only way for a detection signature to have a single Bell state in a correlation class is for the joint-particle kets in that correlation class to add to a scalar multiple of the Bell state, which requires the signature to have all three joint-particle kets in that correlation class. Similarly, a single joint-particle ket in a correlation class is a superposition of all three Bell states in its correlation class. Lastly, it is possible for a state made up of two Bell states in a correlation class to be made up of only two joint-particle kets.  With this information, we are ready to eliminate specific types of detector modes by considering detection signatures containing those modes that must exist in a complete apparatus.

\subsection{No Single-Channel Detector Modes}

\begin{figure}[tb]
	\centering
	\includegraphics[scale=0.35]{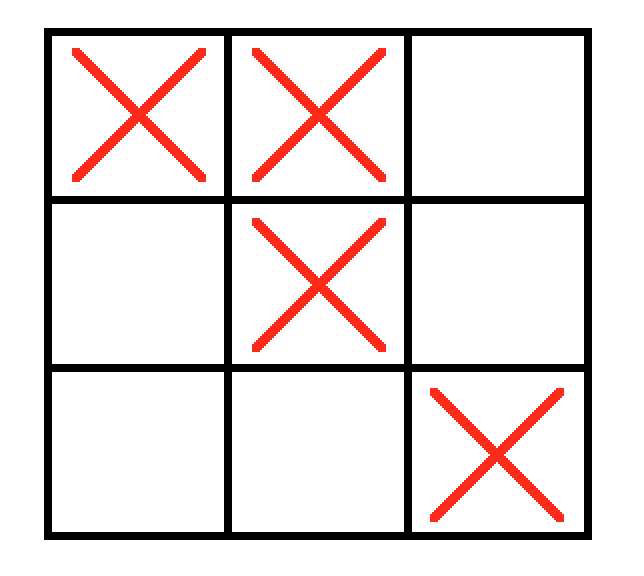}
	\caption{The tic-tac-toe diagram of set A: $\{\ket{\Psi_0^0},\ket{\Psi_0^1},\ket{\Psi_1^1},\ket{\Psi_2^2}\}$. We show that single channel detectors cannot be part of an apparatus that distinguishes these states.}
	\label{fig:00011122}
\end{figure}

We first show that detector modes drawn from a single input channel ($L$ or $R$) cannot exist in an apparatus to distinguish any set of tic-tac-toe winners, by showing that single-channel detectors cannot exist in an apparatus to distinguish one particular set of tic-tac-toe winners.  In particular, we demonstrate that single channel detectors cannot be in an apparatus that can distinguish set A, composed of $\{\ket{\Psi_0^0},\ket{\Psi_0^1},\ket{\Psi_1^1},\ket{\Psi_2^2}\}$ and shown in Figure \ref{fig:00011122}.

Without loss of generality, we will consider single-channel modes in the left channel. Single-channel modes can be superpositions of either one, two, or all three single-particle basis states (referred to casually as `kets' from now on) from that single channel. A single-channel mode with one ket would have the form
\begin{equation}
\ket{S_1}=\ket{a,L},
\label{eq:s1}
\end{equation}
where $a$ equals 0, 1, or 2. We consider this mode in conjunction with a second, arbitrary mode with general form
\begin{equation}
\ket{i}=x\ket{a,R}+y\ket{b,R}+z\ket{c,R}+\ket{L},
\label{eq:arbmode}
\end{equation}
where $b$ and $c$ along with $a$ form a cyclic permutation of 0, 1, and 2, any of the coefficients $x$, $y$ and $z$ could be zero, and $\ket{L}$ is some superposition of kets in the left channel. We can then write all detection signatures involving a click in the 1-ket single channel detector as
\begin{equation}
\ket{S_1}\ket{i}=x\ket{aa}+y\ket{ab}+z\ket{ac}.
\label{eq:ssignature1}
\end{equation}
If $x\neq0$, then there is only one joint-particle ket in the $c=0$ correlation class, which means that the detection signature contains all three Bell states in the $c=0$ correlation class, so that $\ket{\Psi_0^0}$ and $\ket{\Psi_0^1}$ cannot be distinguished. A complete apparatus must contain at least one output mode with $x\neq0$; thus there cannot be any 1-ket single-channel detector modes.

If a single-channel mode had two kets, it would have the form
\begin{equation}
\ket{S_2}=\alpha\ket{a,L}+\beta\ket{b,L}.
\label{eq:s2}
\end{equation}
Using the arbitrary second mode in Equation \ref{eq:arbmode}, we can write all detection signatures involving a click in the 2-ket single-channel detector as
\begin{equation}
\begin{split}
\ket{S_2}\ket{i}=x\alpha\ket{aa}+y\beta\ket{bb}+y\alpha\ket{ab}\\
+z\beta\ket{bc}+x\beta\ket{ba}+z\alpha\ket{ac}.
\end{split}
\label{eq:ssignature2}
\end{equation}
If $x$, $y$ and $z$ are all 0, there is no detection signature. If one or two of them is zero, then two correlation classes only have one joint-particle ket in the signature, which means that all six Bell states in those two correlation classes are in the detection signature. Checking our desired four Bell states in Figure \ref{fig:00011122}, we can see that each correlation class (row) has at least one Bell state (X) in it, so this detection signature would render at least two of those Bell states indistinguishable. So in an apparatus with output mode $\ket{S_2}$, all of the remaining detector modes must have either no right channel kets or all of them.
Those that have all of the right channel kets would create detection signatures with $\ket{S_2}$ that would have two joint-particle kets in each correlation class. That means that they would have to contain least two Bell states in each correlation class.  Thus every detection signature of this form must contain either $\ket{\Psi_0^0}$ or $\ket{\Psi_0^1}$.  Since we also know that all nine Bell states appear in some detection signature involving $\ket{S_2}$, some detection signature must contain both $\ket{\Psi_1^1}$ and another Bell state in our set of four (or $\ket{\Psi_2^2}$ and another Bell state in our set of four).  Thus we have ruled out, overall, the possibility of distinguishing our four Bell states with an apparatus that has a 2-ket single channel detector mode.

If a single-channel mode were to have three kets, it would have the form
\begin{equation}
\ket{S_3}=\alpha\ket{a,L}+\beta\ket{b,L}+\gamma\ket{c,L}.
\label{eq:s3}
\end{equation}
Using the arbitrary mode in Equation \ref{eq:arbmode}, we can write all detection signatures involving a click in the 3-ket single-channel detector as
\begin{equation}
\begin{split}
\ket{S_3}\ket{i}=x\alpha\ket{aa}+y\beta\ket{bb}+z\gamma\ket{cc}+y\alpha\ket{ab}\\
+z\beta\ket{bc}+x\gamma\ket{ca}+z\alpha\ket{ac}+x\beta\ket{ba}+y\gamma\ket{cb}.
\end{split}
\label{eq:ssignature3}
\end{equation}
As we showed at the end of the previous example, if the $c=0$ correlation class has two or more Bell states, this signature cannot distinguish $\ket{\Psi_1^1}$ or $\ket{\Psi_2^2}$ from the other two Bell states in our set. So there must be at least two distinct detection signatures, one that detects $\ket{\Psi_1^1}$ and one that detects $\ket{\Psi_2^2}$ that have $\ket{\Psi_0^2}$ as the only Bell state in the $c=0$ correlation class. This last requirement by itself fixes $x$, $y$ and $z$ up to an overall phase, allowing only one detection signature where we needed at least two. Thus there can be no 3-ket single channel signatures.

Through a series of special cases we have now shown that an apparatus that can distinguish $\{\ket{\Psi_0^0},\ket{\Psi_0^1},\ket{\Psi_1^1},\ket{\Psi_2^2}\}$ may not contain any single-channel output modes. Again, because transformations between the tic-tac-toe winners do not turn single-channel detector modes into multi-channel ones or vice versa, this implies that there are no single channel modes in an apparatus to detect any tic-tac-toe winner.

\subsection{No Detector Modes Have 4 Single-Particle Basis States}
 
Now we show that 4-ket detection modes are banned from an apparatus to distinguish any tic-tac-toe winning set of 4 Bell states. A 4-ket detection mode either has one ket in one input (L/R) channel and three in the other, or has two kets in each input channel. If a 4-ket mode has one ket in one input channel and three in the other, then without loss of generality, let it have one ket in the left channel and three in the right. Let $a$, $b$ and $c$ represent 0, 1 and 2 in an arbitrary cyclic permutation and let $\alpha$, $\beta$, $\gamma$ and $\delta$ be arbitrary nonzero coefficients. Then this detection mode has the form
\begin{equation}
\ket{\text{4-ket}_1}=\alpha\ket{a,L}+\beta\ket{a,R}+\gamma\ket{b,R}+\delta\ket{c,R}.
\label{eq:4ketmode1}
\end{equation}
The detection signature corresponding to two clicks in this detector is
\begin{equation}
\ket{\text{4-ket}_1}\ket{\text{4-ket}_1}=\alpha\beta\ket{aa}+\alpha\gamma\ket{ab}+\alpha\delta\ket{ac}.
\label{eq:4ketsignature1}
\end{equation}
We notice that this is a superposition of three joint-particle kets from different correlation classes. Because one joint-particle ket in a correlation class must be made of a superposition of all three Bell states in that class, this detection signature will be a superposition of all nine Bell states, so it would not be able to distinguish any Bell states. So to distinguish a set of four Bell states, we cannot have 4-ket modes with one ket in one channel and three kets in the other. 

If there are two kets in each channel, there are two possibilities. Either the kets in both channels have the same variable values or they only share one variable value. If they have the same variable values, the detection mode has the form
\begin{equation}
\ket{\text{4-ket}_2}=\alpha\ket{a,L}+\beta\ket{b,L}+\gamma\ket{a,R}+\delta\ket{b,R}
\label{eq:4ketmode2}
\end{equation}
and the detection signature corresponding to two clicks in this detector would be
\begin{equation}
\begin{split}
\ket{\text{4-ket}_2}\ket{\text{4-ket}_2}&=\alpha\gamma\ket{aa}+\beta\delta\ket{bb}\\
&+\alpha\delta\ket{ab}+\beta\gamma\ket{ba}.
\end{split}
\label{eq:4ketsignature2}
\end{equation}
Because $a$ and $b$ are different, $\ket{ab}$ and $\ket{ba}$ are single joint-particle kets in the $c=1$ and $c=2$ correlation classes. Again, these must both be made up of all three Bell states in their correlation classes. $\ket{aa}$ and $\ket{bb}$ are both from the $c=0$ correlation class, and two joint-particle kets are made up of at least two Bell states, so this must have at least two Bell states in the $c=0$ correlation class. In total, this signature must be a superposition of at least eight out of the nine Bell states, so it could not be part of an apparatus to distinguish any four of them.

If there are two kets in each channel and the kets in both channels only share one variable value, then the mode must have the form
\begin{equation}
\ket{\text{4-ket}_3}=\alpha\ket{a,L}+\beta\ket{b,L}+\gamma\ket{a,R}+\delta\ket{c,R}
\label{eq:4ketmode3}
\end{equation}
and the detection signature corresponding to two clicks in this detector would be
\begin{equation}
\begin{split}
\ket{\text{4-ket}_3}\ket{\text{4-ket}_3}&=\alpha\gamma\ket{aa}+\beta\delta\ket{bc}\\
&+\alpha\delta\ket{ac}+\beta\gamma\ket{ba}.
\end{split}
\label{4ketsignature3}
\end{equation}
Here, $\ket{aa}$ and $\ket{bc}$ are the only joint-particle kets in their correlation classes, so again we get all six Bell states from both of those correlation classes. $\ket{ac}$ and $\ket{ba}$ are both in the third correlation class, which gives at least two more Bell states. So again, we get at least eight or nine Bell states, which is not allowed. We cannot have two kets in each channel either, which means 4-ket modes cannot exist in an apparatus to distinguish four of nine Bell states.

\subsection{Distinguishing Set B: At Least One 6-Ket Mode}

\begin{figure}[tb]
	\centering
	\includegraphics[scale=0.35]{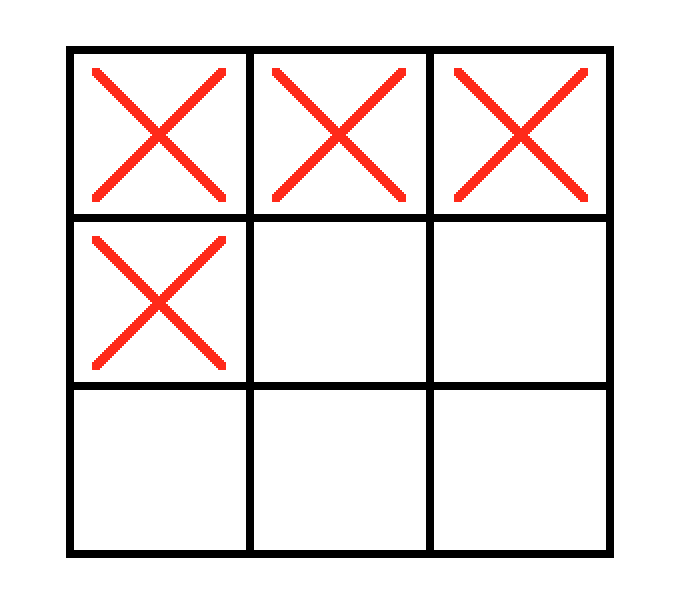}
	\caption{The tic-tac-toe diagram of set B: $\{\ket{\Psi_0^0},\ket{\Psi_0^1},\ket{\Psi_0^2},\ket{\Psi_1^0}\}$. We show that any apparatus that distinguishes these states must have a detector with all six joint-particle kets.}
	\label{fig:00010210}
\end{figure}

We again examine a specific tic-tac-toe winner to determine distinguishability of the entire class. The set we will focus on now is $\{\ket{\Psi_0^0},\ket{\Psi_0^1},\ket{\Psi_0^2},\ket{\Psi_1^0}\}$, shown in Figure \ref{fig:00010210}.
The advantage of using this set is that it has all three Bell states in the $c=0$ correlation class. Thus, viable detection signatures would have to contain at most one of the Bell states in the $c=0$ correlation class, so they must have either zero or all three $c=0$ joint-particle kets. 

In any apparatus that could distinguish this set, we can show that there cannot be any 5-ket modes. Such a mode is missing one input ket, assumed without loss of generality to be in the right channel. It would then have the form
\begin{equation}
\begin{split}
\ket{\text{5-ket}}=\alpha\ket{a,L}+\beta\ket{b,L}+\gamma\ket{c,L}\\
+\delta\ket{a,R}+\epsilon\ket{b,R}.
\end{split}
\label{eq:5ketmode}
\end{equation}
The detection signature corresponding to two clicks in this detector is 
\begin{equation}
\begin{split}
\ket{\text{5-ket}}\ket{\text{5-ket}}=\alpha\delta\ket{aa}+\beta\epsilon\ket{bb}+\alpha\epsilon\ket{ab}\\
+\gamma\delta\ket{ca}+\beta\delta\ket{ba}+\gamma\epsilon\ket{cb}.
\end{split}
\label{eq:5ketsignature}
\end{equation}
We see that this has two joint-particle kets from the $c=0$ correlation class, so it is not allowed, ruling out the possibility of 5-ket modes.

Next, we will show that there cannot be any 2-ket modes. We know such a mode cannot be single-channel, so we will consider only multi-channel 2-ket modes. The two kets may have the same variable value or different variable values. A detector with kets of the same variable value can be written as
\begin{equation}
\ket{\text{2-ket}_1}=\alpha\ket{a,L}+\beta\ket{a,R}
\label{eq:2ketmode1}
\end{equation}
and the detection signature corresponding to two clicks in this detector would be
\begin{equation}
\ket{\text{2-ket}_1}\ket{\text{2-ket}_1}=\ket{aa}.
\label{eq:2ketsignature1}
\end{equation}
This has only one joint-particle ket in the $c=0$ class, so it is not allowed. 

A detector with kets of different variable values can be written as
\begin{equation}
\ket{\text{2-ket}_2}=\alpha\ket{a,L}+\beta\ket{b,R}.
\label{eq:2ketmode2}
\end{equation}
In order to have six orthogonal detectors in the apparatus, at least one other detector would have to contain the $\ket{a,R}$ ket, and could be written as
\begin{equation}
\begin{split}
\ket{i}=v\ket{a,L}+w\ket{b,L}+x\ket{c,L}\\
+\eta\ket{a,R}+y\ket{b,R}+z\ket{c,R},
\end{split}
\label{eq:modewar}
\end{equation}
where $v$, $w$, $x$, $y$ and $z$ may be zero, but $\eta$ is nonzero. The detection signature corresponding to clicks in these two detectors would be
\begin{equation}
\begin{split}
\ket{\text{2-ket}_2}\ket{i}=\alpha\eta\ket{aa}+w\beta\ket{bb}+(y\alpha+v\beta)\ket{ab}\\
+z\alpha\ket{ac}+x\beta\ket{cb}.
\end{split}
\label{eq2ketsignature2}
\end{equation}
We see that this has either one (for $w=0$) or two (for $w\neq 0$) joint-particle kets in the $c=0$ correlation class, so this scenario is also not allowed. Thus there can be no 2-ket modes.

Because of all the forms we have eliminated so far, the only possible detector modes for our apparatus are multi-channel 3-ket modes and 6-ket modes. We will show that there must be a 6-ket mode by showing that this apparatus cannot consist of six 3-ket modes. 
If we have one multi-channel 3-ket mode, it must have two kets in one channel and one ket in the other. Without loss of generality, we will let it have two kets in the left channel, and write it as
\begin{equation}
\ket{\text{3-ket}_1}=\alpha\ket{a,L}+\beta\ket{b,L}+\gamma\ket{d,R},
\label{eq:3ketmode1prev}
\end{equation}
where $d=a$, $b$ or $c$. Then the detection signature corresponding to two clicks in this detector will be
\begin{equation}
\ket{\text{3-ket}_1}\ket{\text{3-ket}_1}=\alpha\gamma\ket{ad}+\beta\gamma\ket{bd}.
\label{eq:3ketsignature1}
\end{equation}
If $d=a$ or $b$, then this signature will have one joint-particle ket in the $c=0$ correlation class.  Thus only $d=c$ is allowed, and
\begin{equation}
\ket{\text{3-ket}_1}=\alpha\ket{a,L}+\beta\ket{b,L}+\gamma\ket{c,R}.
\label{eq:3ketmode1}
\end{equation}

Some detector must have $\ket{a,R}$, $\ket{b,R}$ or $\ket{c,L}$.  If it did not have all three, the detection signature $\ket{\text{3-ket}_1}\ket{\text{3-ket}_2}$ would have fewer than three joint-particle kets in the $c=0$ correlation class and thus more than one of the $c=0$ Bell states, which we are attempting to distinguish. Thus there must be a detector containing $\ket{a,R}$, $\ket{b,R}$ \textit{and} $\ket{c,L}$, and since we are considering the case of an apparatus composed entirely of 3-ket modes, that detector mode takes the form
\begin{equation}
\ket{\text{3-ket}_2}=\delta\ket{c,L}+\epsilon\ket{a,R}+\eta\ket{b,R},
\label{eq:3ketmode2}
\end{equation}
where $\delta$, $\epsilon$ and $\eta$ are all nonzero. Then the detection signature $\ket{\text{3-ket}_1}\ket{\text{3-ket}_2}$ is
\begin{equation}
\begin{split}
\ket{\text{3-ket}_1}\ket{\text{3-ket}_2}=\alpha\epsilon\ket{aa}+\beta\eta\ket{bb}+\gamma\delta\ket{cc}\\
+\alpha\eta\ket{ab}+\beta\epsilon\ket{ba}.
\end{split}
\label{eq:3ketsignature2}
\end{equation}
This has all three joint-particle kets in the $c=0$ correlation class.  However, it also has only one joint-particle ket in each of the other correlation classes and so must contain all three Bell states in each of those classes. Since the detection signature contains at least seven Bell states overall, it must contain at least two of the four Bell states that we are trying to distinguish, making it forbidden in our apparatus.  Thus we see that a successful apparatus must contain at least one detector mode with all six input kets.

\subsection{Distinguishing Set B: No 6-Ket Detector Mode}
\label{Subsection:4Indistinguishable}

Armed with this understanding, we return to the necessary distinguishability criteria from the end of Section \ref{Section:NotationandBackground}. For the set $\{\ket{\Psi_0^0},\ket{\Psi_0^1},\ket{\Psi_0^2},\ket{\Psi_1^0}\}$, shown in Figure \ref{fig:00010210}, we will use the following subset of the conditions given by Equation \ref{eq:criteria}:
\begin{subequations}
\begin{equation}
\bra{\Psi_0^0}\hat{c}^\dagger\hat{c}\ket{\Psi_1^0}=0
\end{equation}
\begin{equation}
\bra{\Psi_0^1}\hat{c}^\dagger\hat{c}\ket{\Psi_1^0}=0
\end{equation}
\begin{equation}
\bra{\Psi_0^2}\hat{c}^\dagger\hat{c}\ket{\Psi_1^0}=0
\end{equation}
\label{eq:subsetcriteria}
\end{subequations}
If we now consider the detector mode that must have six kets:
\begin{equation}
\begin{split}
\ket{\text{6-ket}}=\nu_0^*\ket{0,L}+\nu_1^*\ket{1,L}+\nu_2^*\ket{2,L}\\
+\nu_3^*\ket{0,R}+\nu_4^*\ket{1,R}+\nu_5^*\ket{2,R},
\end{split}
\label{eq:6ketmode}
\end{equation}
(all $\nu$'s must be nonzero) then we get the annihilation operator
\begin{equation}
\hat{c}=\nu_0\hat{a}_0+\nu_1\hat{a}_1+\nu_2\hat{a}_2+\nu_3\hat{a}_3+\nu_4\hat{a}_4+\nu_5\hat{a}_5.
\label{eq:6ketannihilationoperator}
\end{equation}
Plugging this into the conditions in Equations \ref{eq:subsetcriteria} gives Equations \ref{eq:appliedcriteria}:
\begin{subequations}
\begin{equation}
\begin{split}
\bra{\Psi_0^0}\hat{c}^\dagger\hat{c}\ket{\Psi_1^0}=&(\nu_4\nu_3^*+\nu_2\nu_0^*)\\
+&(\nu_5\nu_4^*+\nu_0\nu_1^*)\\
+&(\nu_3\nu_5^*+\nu_1\nu_2^*)=0
\end{split}
\end{equation}
\begin{equation}
\begin{split}
\bra{\Psi_0^1}\hat{c}^\dagger\hat{c}\ket{\Psi_1^0}=&(\nu_4\nu_3^*+\nu_2\nu_0^*)\\
+&e^{i\frac{4\pi}{3}}(\nu_5\nu_4^*+\nu_0\nu_1^*)\\
+&e^{i\frac{2\pi}{3}}(\nu_3\nu_5^*+\nu_1\nu_2^*)=0
\end{split}
\end{equation}
\begin{equation}
\begin{split}
\bra{\Psi_0^2}\hat{c}^\dagger\hat{c}\ket{\Psi_1^0}=&(\nu_4\nu_3^*+\nu_2\nu_0^*)\\
+&e^{i\frac{2\pi}{3}}(\nu_5\nu_4^*+\nu_0\nu_1^*)\\
+&e^{i\frac{4\pi}{3}}(\nu_3\nu_5^*+\nu_1\nu_2^*)=0
\end{split}
\end{equation}
\label{eq:appliedcriteria}
\end{subequations}
Using various linear combinations of these, we can get
\begin{subequations}
\begin{equation}
\nu_4\nu_3^*+\nu_2\nu_0^*=0
\end{equation}
\begin{equation}
\nu_5\nu_4^*+\nu_0\nu_1^*=0
\end{equation}
\begin{equation}
\nu_3\nu_5^*+\nu_1\nu_2^*=0.
\end{equation}
\label{eq:appliedcriteria2}
\end{subequations}
We can then rewrite these as
\begin{subequations}
\begin{equation}
\nu_4\nu_3^*=-\nu_2\nu_0^*
\end{equation}
\begin{equation}
\nu_5\nu_4^*=-\nu_0\nu_1^*
\end{equation}
\begin{equation}
\nu_3\nu_5^*=-\nu_1\nu_2^*.
\end{equation}
\label{eq:appliedcriteria3}
\end{subequations}
Multiplying all of these gives
\begin{equation}
\big|\nu_0\big|^2\big|\nu_1\big|^2\big|\nu_2\big|^2=-\big|\nu_3\big|^2\big|\nu_4\big|^2\big|\nu_5\big|^2.
\label{eq:appliedcriteria4}
\end{equation}
Equation \ref{eq:appliedcriteria4} can only be satisfied if both sides are zero. But because $\ket{\text{6-ket}}$ must have all six kets, none of the $\nu$'s can be zero. Thus these conditions cannot be satisfied, and we have a set of necessary but unachievable conditions for distinguishability of a particular set of four qutrit Bell states.

We have now seen that the bosonic qutrit Bell-state set $\{\ket{\Psi_0^0},\ket{\Psi_0^1},\ket{\Psi_0^2},\ket{\Psi_1^0}\}$ cannot be distinguished with a projective LELM apparatus. Because the set $\{\ket{\Psi_0^0},\ket{\Psi_0^1},\ket{\Psi_0^2},\ket{\Psi_1^0}\}$ is a tic-tac-toe winner, all of the bosonic tic-tac-toe winners must be indistinguishable with an LELM apparatus.

\bibliography{qubitqutritprojnonproj-bibliography}

\bibliographystyle{apsrev}

\end{document}